\documentclass[journal]{IEEEtran}

\usepackage{amsmath}
\usepackage{amssymb}
\usepackage{amsthm}
\usepackage{slashbox,pict2e}
\usepackage{xcolor}
\usepackage{float}
\usepackage{stfloats}
\usepackage{graphicx}
\usepackage{lineno,hyperref}
\modulolinenumbers[5]
\usepackage{epstopdf}
\usepackage{graphicx,cite,multirow,rotating}
\usepackage{setspace}
\usepackage{pgfplots} 
\pgfplotsset{compat=1.18}  

\usepackage{enumitem}   

\usepackage[utf8]{inputenc}
\usepackage{tabularx}
\usepackage{booktabs}

\usepackage{varwidth}

\usepackage{amsfonts}
\usepackage{yfonts}
\usepackage{psfrag}
\usepackage{pifont}
\usepackage{amsfonts}
\usepackage{bbm}
\usepackage{dsfont}
\usepackage{amssymb}
\usepackage{array}
\usepackage{diagbox} 

\usepackage{color}
\usepackage{amsmath,stackrel,dsfont}
\usepackage{amssymb,hyperref,stmaryrd}
\usepackage{pifont}
\modulolinenumbers[5]
\usepackage{amsmath}
\usepackage{epstopdf}
\usepackage{graphicx,cite,multirow,rotating}
\usepackage{setspace}
\usepackage[ruled,vlined,linesnumbered]{algorithm2e}

\usepackage{tikz}
\usepackage{amsmath} 
\usetikzlibrary{shapes,arrows}
\usepackage{varwidth}
\usetikzlibrary{shapes.geometric, arrows}
\usetikzlibrary{positioning,arrows}
\usepackage{amssymb}
\usepackage{amsfonts}
\usepackage{yfonts}
\usepackage{psfrag}
\usepackage{pifont}
\usepackage{amsfonts}
\usepackage{bbm}
\usepackage{dsfont}

\usepackage{color}
\usepackage{amsmath,stackrel,dsfont}
\usepackage{amssymb,hyperref,stmaryrd}
\usepackage{pifont}

\newcommand{\RN}[1]{%
	\textup{\uppercase\expandafter{\romannumeral#1}}%
}

\ifCLASSOPTIONcompsoc
\usepackage[caption=false, font=normalsize, labelfont=sf, textfont=sf]{subfig}
\else
\usepackage[caption=false, font=footnotesize]{subfig}
\fi

\begin{document}
	
	\title{Energy-Efficient Federated Learning with Relay-Assisted Aggregation in IIoT Networks}
	
	\author{Hamid~Reza~Hashempour, Mostafa~Nozari,
		Gilberto~Berardinelli, \textit{Senior Member, IEEE}, Yanjiao Li, \textit{Member, IEEE}, Jie Zhang, \textit{Senior Member, IEEE},		
		Hien Quoc Ngo, \textit{Fellow, IEEE}, and Shashi~Raj~Pandey, \textit{Member, IEEE}
		\thanks{
			
			Hamid~Reza~Hashempour, Jie Zhang, and Hien Quoc Ngo are with the School of Electronics, Electrical Engineering
			and Computer Science, Queen’s University Belfast, Belfast BT7 1NN, UK (email: \{h.hashempoor, jie.zhang, hien.ngo\}@qub.ac.uk).  H.~Q.~Ngo is also with the Department of Electronic Engineering, Kyung Hee University, Yongin-si, Gyeonggi-do 17104, Republic of Korea.
			
			Mostafa~Nozari, Gilberto~Berardinelli and Shashi~Raj~Pandey are with the Department of Electronic Systems, Aalborg University, Aalborg, Denmark
			(e-mails: \{mnozari, gb, srp\}@es.aau.dk). 

            Yanjiao Li is with the Institute of Engineering Technology, University of Science and Technology Beijing, 100083 Beijing, China, and also with the School of Electronic and Electrical Engineering, University of Leeds, LS2 9JT Leeds, U.K. (e-mail: yanjiaoli@ustb.edu.cn).
			
			The work of H.~Q.~Ngo and H. R. Hashempour was supported by a research grant from the Department for the Economy Northern Ireland under the US-Ireland R\&D Partnership Programme. The work of M. Nozari and G. Berardinelli was supported by the Independent Research Fund Denmark under Grant 3105-00077B, and the work of S. R. Pandey was supported by the ``6G-XCEL'' project under Grant 101139194. (\textit{Corresponding authors: Yanjiao Li and Hien Quoc Ngo.})
	} .}


	\maketitle
	
	
	\begin{abstract}
		This paper presents an energy-efficient transmission framework for federated learning (FL) in industrial Internet of Things (IIoT) environments  with strict latency and energy constraints. Machinery subnetworks (SNs) collaboratively train a global model by uploading local updates to an edge server (ES), either directly or via neighboring SNs acting as decode-and-forward relays. To enhance communication efficiency, relays perform partial aggregation before forwarding the models to the ES, significantly reducing overhead and training latency.
		We analyze the convergence behavior of this relay-assisted FL scheme. To address the inherent energy efficiency (EE) challenges, we decompose the original non-convex optimization problem into sub-problems addressing computation and communication energy separately. An SN grouping algorithm categorizes devices into single-hop and two-hop transmitters based on latency minimization, followed by a relay selection mechanism. To improve FL reliability, we further maximize the number of SNs that meet the roundwise delay constraint, promoting broader participation and improved convergence stability under practical IIoT data distributions.
		Transmit power levels are then optimized to maximize EE, and a sequential parametric convex approximation (SPCA) method is proposed for joint configuration of system parameters.
		We further extend the EE formulation to the imperfect channel state information (ICSI).		
		 Simulation results demonstrate that the proposed framework significantly enhances convergence speed, reduces outage probability from $10^{-2}$ in single-hop  to $10^{-6}$ and achieves substantial energy savings, with the SPCA approach reducing energy consumption by at least $2\times$ compared to unaggregated cooperation and up to $6\times$ over single-hop transmission.
	\end{abstract}
	
	
	
	\IEEEpeerreviewmaketitle
	
	\section{Introduction}\label{intro}
	\IEEEPARstart{F}ederated learning (FL) has emerged as a promising privacy-preserving paradigm for collaborative artificial intelligence (AI), particularly suited for the industrial Internet of Things (IIoT). By enabling distributed training of AI models at the network edge without centralized data collection, FL addresses key challenges related to data privacy, communication overhead, and scalability in IIoT systems \cite{T.Zhang,H.Chen,Nguyen, Sisinni, Qu, Hao}. This decentralized approach not only safeguards sensitive industrial data but also accelerates decision-making by aggregating model updates locally, avoiding delays caused by raw data transmission \cite{Zaw}.
	In a typical FL workflow, a central server periodically selects a subset of clients to participate in training. These clients update the model locally based on their data and send the updates to the server, which aggregates them and repeats the process until convergence. This makes FL well-suited for scalable, privacy-aware AI deployment across distributed and heterogeneous IIoT environments \cite{Zhao, Zhan}.

	Short-range, low-power in-X subnetworks (SNs) are being actively explored by both industry and academia for deployment in industrial environments such as robots, sensors, and production modules, aiming to replace traditional wired control infrastructure in the IIoT \cite{Gilberto}. Each SN typically includes a controller access point (AP) and multiple actuators and sensors that communicate locally to collaboratively train a model. For example, a camera integrated into a production module can locally train a model to classify products—representing a typical use case of localized learning within an SN.
	
	In a typical IIoT deployment, tens of such machinery devices operate as independent SNs, each capable of interacting with an edge server (ES). By leveraging FL, these SNs can collaboratively train a global model without sharing raw data, preserving privacy and reducing communication overhead. 
    However, implementing FL in IIoT SNs presents unique challenges. Signal blockages, severe multipath fading in metallic environments, and device mobility can lead to unreliable links and increased communication delays. These delays result in asynchronous or outdated model updates across SNs, introducing model inconsistencies that degrade convergence and learning accuracy. Such effects conflict with the ultra-low latency and high reliability requirements of industrial applications.
    Addressing these challenges is critical for ensuring the timely and efficient operation of FL in IIoT SNs.

	Relay-assisted communication has proven to be a vital enabler for enhancing coverage, reliability, and energy efficiency (EE) in wireless networks \cite{relay1}. In the context of FL, relays can mitigate communication bottlenecks by forwarding aggregated local models to the server, reducing both overhead and latency. 
    This motivates the integration of relay-assisted communication into FL for industrial networks, aiming to reduce latency and improve reliability. To avoid the additional cost of deploying dedicated relay devices, we propose utilizing existing SNs as relays when they have strong communication links to the ES. This approach reduces transmission delay and enhances system reliability. Furthermore, the concept of relay-assisted communication has been explored in the literature.
 For example \cite{Zhang} proposes a rate-splitting-based solution for the Internet of Vehicles, where relay vehicles forward data within a platoon, thereby enhancing communication efficiency.
	Despite these advantages, deploying relay-assisted FL in battery-powered IIoT devices presents significant challenges. These devices must balance the energy demands of local training with the transmission energy required to upload model updates to the server, necessitating efficient resource allocation. To address these issues, \cite{Chen} formulates an energy-efficient FL framework for heterogeneous devices by jointly optimizing weight quantization and wireless transmission, minimizing total energy consumption while ensuring latency and performance requirements. Similarly, \cite{Zhao-Energy} designs an energy-efficient FL scheme for cell-free IoT networks by optimizing total energy consumption under latency constraints. 
	
	Other studies explore diverse strategies to enhance EE in FL. \cite{Pham} leverages unmanned aerial vehicles and wireless-powered communication to minimize the energy consumption of both aerial servers and users in FL networks. \cite{Energy-RIS} formulates an energy minimization problem for reconfigurable intelligent surface (RIS)-assisted FL systems while adhering to training time constraints. \cite{Salh} investigates joint optimization of bandwidth allocation, CPU frequency, transmission power, and learning accuracy to minimize energy consumption while meeting FL time requirements. Additionally, \cite{Abiad} introduces a resource allocation scheme for non-orthogonal multiple access (NOMA)-enabled, relay-assisted IoT networks to reduce overall energy consumption in FL systems.
	
        	However, none of the aforementioned works specifically address both EE and cooperative communication in the context of industrial SNs, which require high reliability and low latency. Our recent work in \cite{Hashempour} investigates power-efficient algorithms for cooperative communication in IIoT systems, but it focuses solely on communication within SNs and does not consider FL. Building on this foundation, the present paper introduces a novel, energy-efficient, relay-assisted transmission protocol tailored for FL in IIoT networks. 
        	In addition to the perfect channel state information (PCSI) scenario, we further incorporate practical minimum mean-square error (MMSE)-based imperfect CSI (ICSI) into the communication model and develop an effective-SNR expression that explicitly accounts for estimation uncertainty. This enables a unified EE optimization across both CSI regimes within the same SPCA framework.
            Compared with existing works, the key distinctions of our approach are threefold: (i) the use of existing SNs as relays without requiring dedicated infrastructure, (ii) the joint consideration of FL and cooperative communication under both PCSI and ICSI conditions, and (iii) a unified EE optimization framework based on SPCA that accounts for practical channel estimation errors.
	The proposed protocol is designed to satisfy stringent latency and energy constraints by classifying SNs into single-hop and two-hop transmission modes based on their CSI. In the two-hop mode, relay nodes perform partial aggregation of local models before forwarding them to the ES, thereby enhancing resource utilization and reducing overall training latency.
	
	The key contributions of this paper are summarized as follows:
	\begin{itemize}[label=$\bullet$]
		\item We formulate the EE problem for relay-assisted FL under a time division multiple access (TDMA) protocol. Given the interdependence between computational and transmission energy, we decompose the problem into manageable sub-problems to enable efficient optimization.
		
		\item We analyze the outage probability with a focus on transmission delay and evaluate the convergence of the proposed relay-assisted FL framework compared to standard single-hop FL under non-IID data distributions. The analysis highlights the advantages of the relay-assisted approach in achieving faster convergence and greater resilience to data heterogeneity.
		
		\item Building on our previous work in \cite{Hashempour}, we propose an algorithm to classify SNs into single-hop and two-hop transmission groups and select optimal relays, aiming to minimize transmission delay while satisfying strict latency constraints.
		
		\item To tackle the non-convexity of the EE problem, we adopt a sequential parametric convex approximation (SPCA) method to jointly optimize system parameters, including transmit power and relay selection.

			\item We extend our framework by incorporating MMSE-based channel estimation into the FL pipeline, deriving the effective SNR expression under ICSI and modifying the SPCA algorithm accordingly to handle channel estimation error effect.
			
			\item Simulations demonstrate substantial gains in EE, reduced outage probability, and faster FL convergence compared with classical single-hop and two-hop schemes. Under both PCSI and ICSI, the proposed method consistently yields the lowest uplink energy for a fixed FL accuracy target, with ICSI performance approaching PCSI as the pilot length increases.

	\end{itemize}
	
	To the best of our knowledge, this is the first work to integrate relay-assisted TDMA transmission with stringent timing constraints for FL training and derive an energy-efficient solution within industrial IIoT networks.
	
	The remainder of this paper is structured as follows: Section~\ref{Sys_Model} introduces the system model and the proposed communication protocol. Section~\ref{Optim} presents the optimization methodology for EE. Section~\ref{Convergence} analyzes the convergence of the relay-assisted FL framework. Numerical results are discussed in Section~\ref{Simulat}, and the conclusions are summarized in Section~\ref{conc}.

	\textit{Notation}: We use bold lowercase letters for vectors and bold uppercase letters for matrices. The notation $(\cdot)^T$ and $(\cdot)^H$  denote the transpose operator and the conjugate transpose operator, respectively.  $\triangleq$ denotes a definition. $\mathbb{R}^{N \times 1}$ and $\mathbb{C}^{N \times 1}$ denote the sets of $N$-dimensional real and complex vectors, respectively.  $\mathbb{C}^{M \times N}$ stands for the set of $M \times N$ complex matrices.   $\mathrm{diag}\{\cdot\}$ constructs a diagonal matrix from its vector argument. The $\mathrm{Exp}(\lambda)$ distribution represents the exponential distribution with $\lambda$ as the rate parameter. Similarly, the $\mathcal{G}(\mu, \sigma^2)$ distribution represents the Gaussian distribution with mean $\mu$ and variance $\sigma^2$. $\mathbb{E}\{\cdot\}$ denotes the expectation operator, which represents the expected value of a random variable.

	Table \ref{Table-1} presents the main parameters and variables associated with this study to enhance the readability of the paper. 
	\begin{table}
		\small
		\renewcommand{\arraystretch}{1.3}
		\caption{Key notations used in this paper.}
		\centering
		\label{Table-1}
		\resizebox{\columnwidth}{!}{
			\begin{tabular}{|c|p{60mm}|}
				\hline
				\textbf{Notation}  &  \textbf{Definition} \\
				\hline
				$\mathcal{N}$, $N$, $n$  &  Set, number, and index of SNs. \\
				\hline
				$\mathcal{K}$, $K$, $k$  &  Set, number, and index of relays. \\
				\hline
				$\mathcal{N}_{\text{1h}}$, $\mathcal{N}_{\text{2h}}$; $\mathcal{N}_{\text{2h},k}$ 
				&  The set of SNs scheduled for single-hop/two-hop transmission; SNs associated with relay $k$. \\
				\hline
				$\text{S}_n$, $\text{R}_k$  & The $n$-th SN and $k$-th relay. \\
				\hline   
				$h_n^\text{a}$, $h_{n,k}^\text{r}$	  &  The channel of the $n$-th SN to the ES; channel from the $n$-th SN to the $k$-th relay (\( \text{R}_k \)). \\
				\hline
				$h_k^\text{a}$ & The channel vector between the ES and \( \text{R}_k \). \\
				\hline
				$P_n$ & The transmit power of the $n$-th SN (W). \\
				\hline
				$\lvert B \rvert$, $W$ & Uploaded model size of each SN; available bandwidth (Hz). \\ 
				\hline
				$T$ & Uplink timeslot (s). \\
				\hline
				$\sigma_0$ & AWGN noise power (W). \\
				\hline
				$D_n$, $\lvert D_n \rvert$ & Local trainable dataset; total number of training samples in the dataset of the $n$-th SN. \\
				\hline
				$\lvert D_k^\text{r} \rvert$ & Effective dataset size at the relay \( \text{R}_k \), including contributions from associated SNs and the relay itself. \\
				\hline
				$w_n/v_n$ & The local model of the $n$-th SN before/after training (i.e., $w_n$ is the initial local model and $v_n$ is the updated model). \\
				\hline
				$w_k^{(\text{r})}$ & Aggregated model at relay \( \text{R}_k \), combining contributions from \( \mathcal{N}_{\text{2h},k} \) and its own model. \\
				\hline
				$w$ & Global model aggregated at the ES. \\
				\hline
			\end{tabular}
		}
	\end{table}
	
	\section{System Model}\label{Sys_Model}
\begin{figure}
		\centering
		\includegraphics[width=1\linewidth]{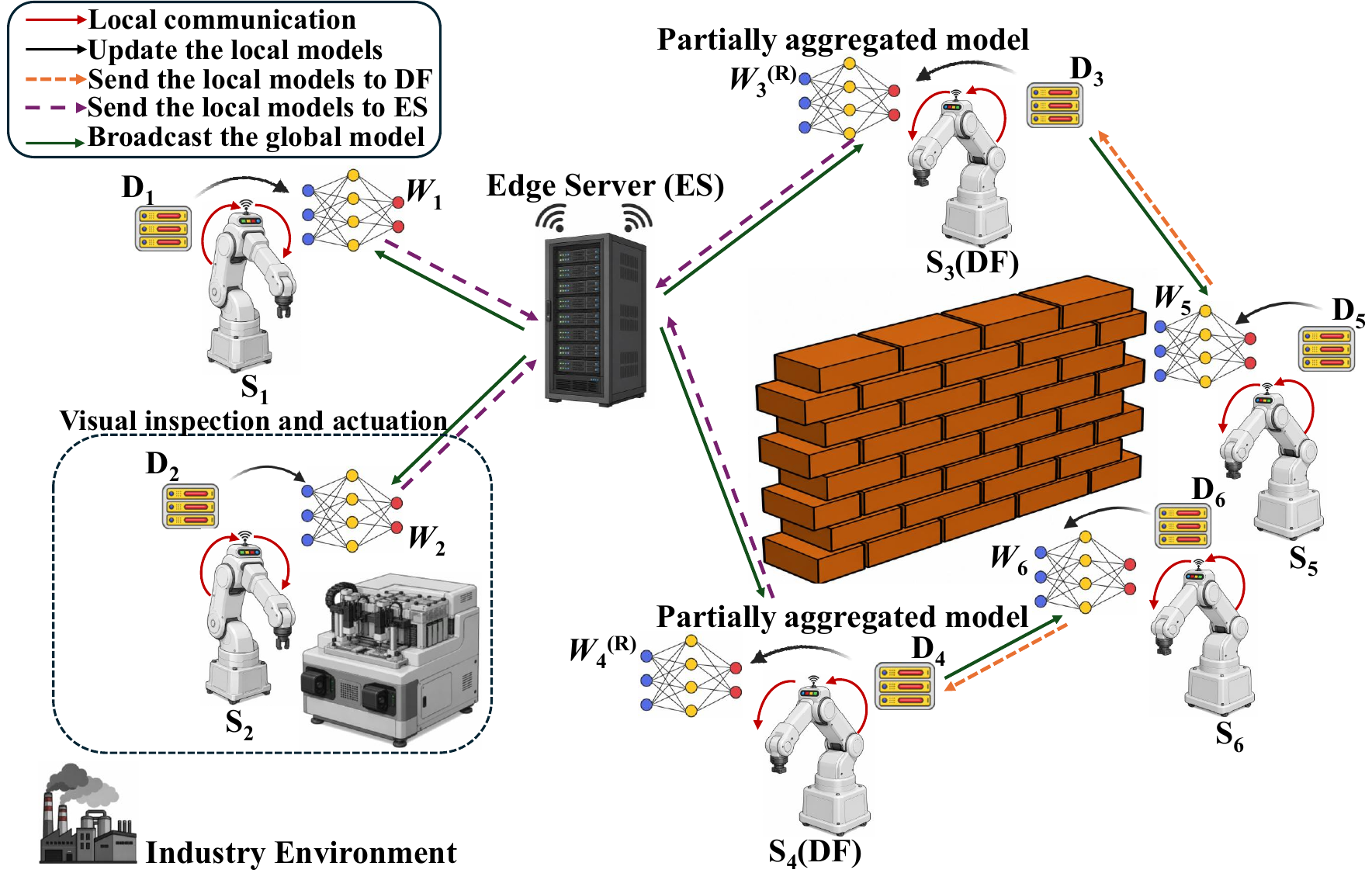}
		\caption{System model for FL with multiple SNs in an IIoT network.}
		\label{fig1a}
	\end{figure}

	We consider an industrial wireless network of $N$ SNs, denoted by $\mathcal{N} = \{1,2, \dots, N\}$. Each SN $\mathrm{S}_n$ operates within a factory cell, consisting of an AP and a set of end devices (e.g., sensors, actuators). The end devices collect raw data from the environment (e.g., images for inspection, quality measurements) and forward it to the AP. 
 A representative application scenario is a smart manufacturing system, where each AP supervises an inspection or monitoring station (e.g., defect detection or predictive maintenance) and enables timely control actions. In such industrial FL systems, EE is a critical performance metric, as SNs are often energy-constrained and frequent model updates incur significant communication overhead. Improving EE is therefore essential to prolong the device lifetime while maintaining low latency and high reliability \cite{Zaw,relay1,Chen,Zhao-Energy,Pham,Energy-RIS,Salh,Abiad,Hashempour,Energy1,Khosravirad}.
 Each AP processes this data and trains a local model $w_n$ using its dataset $D_n$ of size $|D_n|$. To enable collaborative learning across the network, every SN acts as a FL client, submitting its model to a central edge server (ES), as illustrated in Fig.~\ref{fig1a}. To account for variable uplink (UL) channel quality, SNs are partitioned into: (i) $\mathcal{N}_\text{1h}$, which send their updates directly to the ES, and (ii) $\mathcal{N}_\text{2h}$, which utilize two‑hop relaying.

	The relaying is performed by a set of relay APs $\mathcal{K}$. Each relay $R_k \in \mathcal{K}$ serves a subset $\mathcal{N}_\text{2h,k} \subseteq \mathcal{N}_\text{2h}$, collecting and partially aggregating the local model updates from its associated SNs before forwarding the result to the ES. This allows SNs with weaker direct links to participate reliably in the FL process.
	We denote the direct link between $\mathit{S}_n$ and the ES as $h_n^{\rm a}$, the link between $\mathit{S}_n$ and its relay $R_k$ as $h_{n,k}^{\rm r}$, and the link between the relay and the ES as $h_{k}^{\rm a}$. Each SN chooses its UL path based on channel conditions: if $h_n^{\rm a}$ is strong, it sends $w_n$ directly to the ES; otherwise, it adopts the relay path over $h_{n,k}^{\rm r}$, where $R_k$ decodes, aggregates, and then transmits the result to the ES over $h_{k}^{\rm a}$. Upon receiving both direct and relayed updates, the ES applies a global aggregation (e.g., FedAvg) to produce an updated global model $w$, which it then broadcasts back to all SNs. This scheme allows FL to iteratively refine local models, improve decision accuracy, preserve data privacy, and reduce communication overhead across the IIoT network.

	Next, we present the preliminaries necessary for constructing the delay and consumption energy model in our relay-assisted FL framework. This includes an overview of FL model training, the wireless communication model, and the processing and transmission model. These elements collectively form the foundation for deriving the EE model, which is critical for optimizing EE in the proposed relay-assisted FL system.
	\subsection{FL Model}
	We adopt an FL framework in which SNs collaboratively train a global model \( w \) without sharing raw data. The workflow is as follows:
	1. Each SN trains a local model \( w_n \) on its dataset \( D_n \).
	2. SNs in \( \mathcal{N}_{\text{1h}} \) transmit their models directly to the ES.
	3. SNs in \( \mathcal{N}_{\text{2h}} \) send their models to assigned relays; each relay aggregates models from \( \mathcal{N}_{\text{2h},k} \) and forwards the result to the ES.
	4. The ES aggregates all received models to obtain the global model \( w \).
	5. The global model \( w \) is broadcast to all SNs for the next round. The details are as follows:    
	In each round, SNs train a statistical model locally on their datasets. The objective of SN \( \text{S}_n \) is to minimize the local training loss over its dataset \( D_n \), given by \cite{Hao}
	\begin{align}\label{FL-model}
		& \displaystyle  \min_{w_n} F_n(w_n) = \sum_{j \in D_n} \frac{1}{|D_n|}l(w_n, x_j),
	\end{align}
	where \( l(w_n, x_j) \) denotes the loss function for a training sample \( x_j \).
	Each SN uses stochastic gradient descent (SGD) to optimize its local model over \( e \) epochs. The updated model \( w_n \) is then transmitted to the ES or to its assigned relay.
	Specifically, in each round, the $n$th SN computes the training loss and then updates the weights using gradient descent as
	\begin{align}
		v_n &  =w_n -\eta \nabla  F_n(w_n),
	\end{align}
	where $v_n$ represents the updated model parameter of the $n$th SN, and $\eta$ denotes the learning rate.
	\subsubsection{Relay Aggregation}
	In the two-hop communication scenario, after receiving
	and decoding the local models, relay nodes aggregate the local models received from their associated SNs along with their own models. 
	Without loss of generality, the well-known FedAvg \cite{McMahan} is adopted
	in this work to aggregate the local trained models. Thus, the aggregated model \( w_k^{(\text{r})} \) at relay \( \text{R}_k \) is computed as
	\begin{align}\label{partially_aggreg}
		w_k^{(\text{r})} = \frac{\sum_{n \in \mathcal{N}_{\text{2h},k}} |D_n| v_n + |D_k| v_k}{|D_k^\text{r}|},
	\end{align}
	where $v_k$ denotes the updated model at relay 
	$\text{R}_k$, $ |D_k^\text{r}| = \sum_{n \in \mathcal{N}_{\text{2h},k}} |D_n| + |D_k|$ is the effective dataset size at relay \( \text{R}_k \), including contributions from \( \mathcal{N}_{\text{2h},k} \) and its own dataset.
	In this paper, we assume synchronized aggregation at relay \( \text{R}_k \)  to reduce communication overhead and prevent model staleness. We also assume that all local models are perfectly decoded at the relay without error.
	To address delays, we set a latency threshold to drop slow SNs and optimize resource allocation to minimize waiting time.
	\subsubsection{Global Model Aggregation}
	The ES aggregates models from both single-hop SNs (\( \mathcal{N}_{\text{1h}} \)) and relays (\( \mathcal{K} \)). The global model \( w \) is computed as
	\begin{align}\label{global_model}
		w = \frac{\sum_{n \in \mathcal{N}_{\text{1h}}} |D_n| v_n + \sum_{k \in \mathcal{K}} |D_k^\text{r}| w_k^{(\text{r})}}{\sum_{n \in \mathcal{N}_{\text{1h}}} |D_n| + \sum_{k \in \mathcal{K}} |D_k^\text{r}|}.
	\end{align}

	\subsection{Wireless Communication Model}
	Fig.~\ref{fig1a} illustrates an example of an IIoT network comprising six SNs and an ES. 
	Among these SNs, S5 and S6 operate in a two-hop cooperative communication mode because their direct links to the ES are obstructed. In this scenario, S3 and S4—although they normally maintain direct single-hop links to the ES—also serve as decode-and-forward (DF) relays to assist S5 and S6 in forwarding their transmissions. In contrast to S5 and S6, S1 and S2 communicate directly with the ES via single-hop links.
	The ES and all SNs are assumed to be time-synchronized and operate within a shared frequency band. To mitigate interference, the ES centrally allocates time resources to SNs using a TDMA scheme. We assume that the ES has global knowledge of the CSI for all communication links: (i) between each SN’s AP and the ES, and (ii) between APs of different SNs, which are potential relay pairs. The extension to the ICSI case is discussed in Section~\ref{ICSI-discuss}.
	This CSI is obtained during an initial training phase in which each AP transmits reference signals for channel estimation. 
	In dynamic network environments, the channel responses may vary over time; however, they are assumed to remain quasi-static within each FL training round.
	Thus, the channel training procedure can be performed periodically with a relaxed update frequency to accommodate practical deployment conditions.
	The ES scheduler utilizes the complete CSI from all SN links to optimize communication parameters. Specifically, it determines the appropriate SNs for single-hop and two-hop transmissions, allocates transmission rates, manages time resource distribution, and configures transmit power for energy-efficient communication.

\begin{figure} 
		\centering
		\includegraphics[width=1\linewidth]{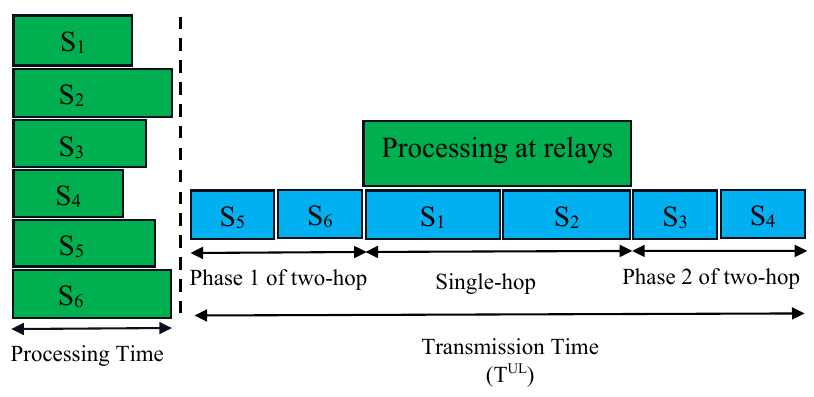}
		\caption{Proposed implementation framework for an FL algorithm using the cooperative TDMA protocol.}
		\label{time}
	\end{figure}

	The proposed protocol for achieving EE in relay-assisted FL is illustrated in Fig.~\ref{time}. Each communication round is divided into two main phases: the local processing phase and the UL transmission phase.\footnote{We omit the downlink (DL) transmission time required to share the global model with all SNs, based on the assumption that the ES has ample power and computational resources. This simplification does not affect generality, as the model can be easily extended to include both UL and DL phases.}
	The processing phase duration depends on the computation
	capacity of each SN and may vary across the network. However, to ensure synchronization, the overall processing time is determined by the slowest SN in the network.
	
	The UL transmission phase is further divided into three variable-length sub-slots: (i) the first phase of two-hop transmissions, (ii) single-hop transmissions, and (iii) the second phase of two-hop transmissions. SNs belonging to the set \( \mathcal{N}_{\text{2h}} \) utilize the first sub-slot to transmit their local models to their assigned relay SNs. Each SN in \( \mathcal{N}_{\text{2h}} \) is served by a relay SN that offers the most favorable channel conditions.
	Once the relay receives the models from its associated SNs, it decodes the packets, aggregates the received models together with its own local model, and forms a partially aggregated model. During the second-phase sub-slot, this aggregated model is forwarded to the ES. Meanwhile, SNs assigned to the single-hop set \( \mathcal{N}_{\text{1h}} \) transmit their local models directly to the ES during the single-hop sub-slot.

	Below, we outline the signal model for single-hop and two-hop DF relaying.
	\subsubsection{Single-hop Transmission}
	SNs in \( \mathcal{N}_{\text{1h}} \) directly transmit their local models \( w_n \) to the ES over wireless channels \( h_n^\text{a} \). The signal-to-noise ratio (SNR) of $\text{S}_n$ and the UL transmission rate, respectively, are
	\begin{align} 
		g_{n}^\text{d}&=\frac{P_n |h_n^\text{a}|^2}{\sigma_0},\ \ \forall  n \in \mathcal{N}_{\text{1h}},
	\end{align}
	\begin{align}
		r_{n}^\text{d}  =  \log_2\left(1 + g_{n}^\text{d}\right),\ \ \forall  n \in \mathcal{N}_{\text{1h}},
	\end{align}
	where \( P_n \) is the transmit power of \( \text{S}_n \), and \( \sigma_0 \) is the AWGN noise power and \( W \) is the available bandwidth.
	
	\subsubsection{Two-hop Transmission}
	The SNs with strong link to ES are categorized as potential relays and the set of all potential relays are $\mathcal{K} \subseteq \mathcal{N}$. Thus, 
	SNs in \( \mathcal{N}_{\text{2h},k} \) first transmit their local models to relay \( \text{R}_k \) over channels \( h_{n,k}^\text{r} \). The relay aggregates the received models, and then transmits the aggregated model \( w_k^{(\text{r})} \) to the ES over channel \( h_k^\text{a} \). The SNR and rates for transmission in the first phase are respectively, given by
	\begin{align}
		g_{n,k}&=\frac{P_n |h_{n,k}^\text{r}|^2}{\sigma_0},\ \ \forall  n \in \mathcal{N}_{\text{2h}}, \ \forall  k \in \mathcal{K},  \\
		r^{(1)}_{n,k} &=  \log_2\left(1 + g_{n,k}\right),\ \ \forall  n \in \mathcal{N}_{\text{2h}}, \ \forall  k \in \mathcal{K}, 
	\end{align}
	where  the superscript ${(1)}$ indicates the first phase in the two-hop cooperative transmission.
	After aggregating the model and transmitting, the SNR and the achievable rate from the $k$th relay to the ES  are respectively given by
	\begin{align}\label{eq7}
		g_{k}&=\frac{P_k |h_k^\text{a}|^2}{\sigma_0}, \ \forall  k \in \mathcal{K} ,\\
		r^{(2)}_{k} &= \mathrm{log}_2 \left( 1+ g_{k} \right),\  \ \forall  k \in \mathcal{K} ,
	\end{align}
	where the superscript ${(2)}$ denotes the second phase in the two-hop cooperative transmission and \( P_k \) is the transmit power of relay \( \text{R}_k \).

	\subsection{Processing and Transmission Model}
	
	The FL process involving SNs and their serving ES is illustrated in Fig.~\ref{fig1a}. Each iteration of this FL process comprises three distinct stages: (i) \textit{local computation}, where each SN independently computes its local FL model through multiple local iterations using its own dataset and the most recent global model received from the ES; (ii) \textit{model transmission}, wherein SNs transmit their locally computed models either directly to the ES or indirectly through relay nodes---these relay nodes perform partial aggregation by combining their local models with all models received from other SNs before forwarding the aggregated result to the ES; and (iii) \textit{global aggregation and broadcast}, in which the ES aggregates all received models to generate an updated global FL model and subsequently broadcasts this updated model back to the SNs.

	\subsubsection{Local Computation}
	Let $f_n$ denote the computation capacity of $\text{S}_n$, quantified by the number of CPU cycles per second. The computation time required at $\text{S}_n$ for data processing is given by
	\begin{align}\label{tau}
		\tau_n = \dfrac{I_n C_n D_n}{f_n},\ \forall n \in \mathcal{N}.
	\end{align}
	Here, $C_n$ (cycles/sample) denotes the number of CPU cycles required to compute a single data sample at the AP of $\text{S}_n$. The parameter $I_n$ represents the number of local training iterations performed by $\text{S}_n$ before global aggregation. The value of $I_n$ can affect both model accuracy and resource consumption, and is typically determined based on system-level trade-offs between communication and computation efficiency. While the precise calculation of $I_n$ can be adapted based on desired accuracy or energy constraints—as discussed in prior work such as \cite{Energy1}—we do not delve into its derivation here to maintain simplicity and avoid redundancy.
	As per \cite{Energy1}, the energy consumption for performing a total of $C_n D_n$ CPU cycles at $\text{S}_n$ is
	\begin{align}
		E_{n1}^L = \kappa  C_n D_n f_n^2,
	\end{align}
	where $\kappa$ denotes the effective switched capacitance dependent on the chip architecture. To compute the local FL model, $\text{S}_n$ must perform $ C_n D_n$ CPU cycles across $I_n$ local iterations, resulting in the total local computation energy at $\text{S}_n$ as
	\begin{align}
		E_{n}^L = I_n E_{n1}^L = \kappa I_n C_n D_n f_n^2.
	\end{align}
	It is important to note that the energy consumed for partially aggregation at the relays is omitted in this analysis for simplicity, as the computation at the relays is non-iterative and incurs negligible energy costs compared to iterative local updates performed by the SNs.
	\footnote{The energy associated with the decoding and re-encoding operations at the relays is negligible compared to the transmission energy, as these operations involve low-power signal processing, whereas transmission requires high-power amplifiers that dominate the overall energy consumption \cite{Xu2025,Maaz2016,Karvonen2004}.}

	\subsubsection{Wireless Transmission}
	In this phase, $\text{S}_n$ must transmit the local FL model to either the ES or the associated relay. We assume that the local FL model has a fixed dimensionality across all SNs, implying that each SN uploads model updates of the same size, denoted by $|B|$.
	Considering time division multiplexing over a bandwidth $W$, a packet of $|B|$ bits for $\text{S}_n$ can be transmitted within $\dfrac{|B|}{W r_{n}^\text{d} }$ time. Transmitting packets from all single-hop SNs in a TDMA manner results in a total transmission time
	\begin{align}
		T_{\textrm{1h}} &= \sum_{n \in \mathcal{N}_{\text{1h}}} \dfrac{|B|}{W r_{n}^\text{d} }.
	\end{align}
	To execute DF, the signal from $\text{S}_n$ must be accurately decoded by the relay $k^*_n \in \mathcal{K}$ with the strongest channel gain, then after receiving
	all the model parameters from the SN set $\mathcal{N}_{2h,k^*_n}$, relay $R_{k^*_n}$
	aggregates the local model according to \eqref{partially_aggreg}. Then, re-encoded into a new message. The duration of over-the-air time required for successful transmission of a packet containing $|B|$ bits by $\text{S}_n$ is therefore
	$\dfrac{|B|}{W   r_{n,k^*_n}^{(1)} }+\dfrac{|B|}{W  r_{k^*_n}^{(2)}},\ \forall  n \in \mathcal{N}_{\text{2h}},\ k^*_n \in \mathcal{K}$.
	Let's denote the total transmission time for all SNs in the first and second phases of the two-hop method as 
	\begin{align}\label{eq10}
		T_{\textrm{2h}}^{(1)} & = \sum_{n \in \mathcal{N}_{\text{2h}}} \dfrac{|B|}{W  r_{n,k^*_n}^{(1)} }, \\
		T_{\textrm{2h}}^{(2)} & = \sum_{k \in \mathcal{K}} \dfrac{|B|}{W  r_k^{(2)}}.
	\end{align}
	The total transmission time in both single-hop and two-hop cases is then calculated as
	\begin{align}\label{eq18}
		T^{\textrm{UL}}&= T_{\textrm{1h}} + T_{\textrm{2h}}^{(1)} +  T_{\textrm{2h}}^{(2)}, 
	\end{align}
	To transmit data of size $|B|$ within a time duration $T^{\textrm{UL}}$, the wireless transmit energy will be given by
	\begin{align}\label{eq-Et}
		E^T & =\dfrac{|B|}{W} \left [\sum_{n \in \mathcal{N}_{\text{1h}}} \dfrac{P_n}{ r_{n}^\text{d} }+\sum_{n \in \mathcal{N}_{\text{2h}}}\dfrac{P_n }{   r_{n,k^*_n}^{(1)} }
		+ \sum_{k \in \mathcal{K}} \dfrac{P_{k}}{ r_k^{(2)}}\right].
	\end{align}
	Considering the FL model outlined above, each user's energy consumption comprises both local computation energy $E_{n}^L$ and wireless transmission energy $E^T$. Let's denote the number of global iterations as $I_0$. Then, the total UL energy consumed by all SNs participating in FL will be
	\begin{align}
		E & = I_0 \left( E^T + \sum_{n \in \mathcal{N}} E_{n}^L  \right).
	\end{align}
	Each SN $S_n$,  where \( n \in \mathcal{N} \), possesses a local dataset used for training. As illustrated in Fig.~\ref{time}, the time required to complete one round of the FL process, denoted by \( T^\mathrm{t} \), consists of two main components: the local computation time and the UL transmission time. Since all SNs must complete their local training before proceeding to the transmission phase, the total computation time is determined by the slowest SN, that is, the maximum computation time among all SNs. According to \eqref{tau}, the total time for each round is given by
	\begin{align}
		T^\mathrm{t} = \max_n (\tau_n) + T^{\mathrm{UL}}.
	\end{align}
	As stated earlier, the downlink transmission time is neglected. Consequently, the total completion time of the FL process is
	\begin{align}
		T^\mathrm{c} = I_0 T^\mathrm{t}.
	\end{align}
	In a factory environment, maintaining strict delay constraints is particularly critical, as each SN must support both low-latency industrial control communications and FL model updates. Let \( T^{\mathrm{th}} \) denote the maximum allowable time for completing one round of the FL process. To ensure timely completion, the sum of the computation and communication times must satisfy the constraint \( T^\textrm{t} \leq T^{\mathrm{th}} \).
	If the total time exceeds the allocated threshold, it may cause buffer overflows or disruptions in critical industrial communications, leading to what we refer to as a \textit{system outage}\cite{Khosravirad}. To mitigate this risk, the ES proactively manages the scheduling process by excluding SNs with weak channel conditions that could cause the round duration to violate the delay constraint. The details will be discussed in the next section.

	\section{Proposed Method for Energy Efficiency}\label{Optim}
	In this section, we establish the EE problem for FL. Given the nonconvex nature of the problem, obtaining the globally optimal solution is challenging. Therefore, we propose a low-complex iterative algorithm to address the energy minimization problem.
	\subsection{Problem Formulation} 
	Our objective is to minimize the total energy consumption of all SNs while adhering to a latency constraint. This energy-efficient optimization problem can be formulated as follows
	\begin{subequations} \label{P0}
		\allowdisplaybreaks
		\begin{align} &\min_{\mathbf{P},\mathbf{f}, \mathcal{N}_{\text{1h}},\mathcal{N}_{\text{2h}},\mathcal{K}} E, \label{P0-a}\\
			\mathrm{s.t.}\ & 
			\left(    \dfrac{I_n C_n D_n}{f_n}   +T^{\textrm{UL}} \right)\leq  T^{\mathrm{th}}, \label{P0-b}
			\\& 0 \leq P_n \leq P_{\max}, \ \forall n \in \mathcal{N},\label{P0-c}\\
			& 
			0 \leq f_{n} \leq f_{n}^{\max}, \ \forall n \in \mathcal{N},\label{P0-e}	
		\end{align} 
	\end{subequations}
	where $\mathbf{P} = [P_1, P_2, \ldots, P_{N}]$ and $\mathbf{f} = [f_1, f_2, \ldots, f_{N}]$ denote the vectors of transmit powers and computational capacities of all SNs, respectively. $P_{\text{max}}$ and $f_n^{\max}$ represent the maximum allowable transmit power and the maximum computational capacity of $\text{S}_n$, respectively. Constraint \eqref{P0-b} addresses the requirement for low latency, while \eqref{P0-c} sets the power limit.

	The formulated EE problem poses a challenge due to its non-convexity and the strong coupling among decision variables. Therefore, we initially decouple the problem into the computing resource management problem and the transmit EE problem.
    
\noindent\textbf{Remark:}
Although the optimization problem in \eqref{P0} is formulated in terms of 
energy and latency, these variables directly determine (i) how many SNs can 
meet the roundwise deadline and thus participate in each FL update, and 
(ii) the timeliness of global aggregation. Both aspects are known to strongly 
influence the convergence behavior of FedAvg, particularly under non-IID data 
distributions \cite{Energy1,li2019convergence}.  
Therefore, improving energy efficiency and reducing communication delay 
indirectly enhance FL convergence without explicitly embedding the FL loss 
function into the optimization objective, which follows standard practice in 
wireless FL optimization.
\subsection{Designing the SN CPU Frequency}\label{CPU Freq}
Initially, we optimize the frequency once the number of training iterations required to achieve a specific accuracy is known. Minimizing the total energy consumption across all SNs is equivalent to minimizing the individual energy consumption of each SN. For each SN, $E_{n}^L$ is an increasing function with respect to $f_n$. The time constraint \eqref{P0-b} of Problem \eqref{P0} implies that each SN should operate at the lowest frequency $f_n^*$ permitted by the delay constraint. Thus, we obtain
	\begin{align}
		f_n^* = \dfrac{I_n C_n D_n}{T^{\mathrm{th}}  -T^{\textrm{UL}} }. 
	\end{align}
	
	\subsection{Proposed Method for Relay Selection and Transmit Power Control}\label{Optim_relay_power}
	After minimizing the computation energy $E_{n}^L$, the formulated problem transforms into the energy transmission efficiency problem, expressed as follows
	\begin{subequations} \label{P1}
		\allowdisplaybreaks
		\begin{align} &\min_{\mathbf{P}, \mathcal{N}_{\text{1h}},\mathcal{N}_{\text{2h}},\mathcal{K}} \left [\sum_{n \in \mathcal{N}_{\text{1h}}} \dfrac{P_n}{ r_{n}^\text{d} }+\sum_{n \in \mathcal{N}_{\text{2h}}}\dfrac{P_n }{   r_{n,k^*_n}^{(1)} }
			+ \sum_{k \in \mathcal{K}} \dfrac{P_{k}}{ r_k^{(2)}}\right], \label{P1-a}\\
			\mathrm{s.t.}\ & 
			T^{\textrm{UL}}\leq  T^{\mathrm{eff}}, \label{P1-b}
			\\& 0 \leq P_n \leq P_{\max}, \ \forall n \in \mathcal{N},\label{P1-c}
		\end{align} 
	\end{subequations}
	where $T^{\mathrm{eff}}$ is the transmission time requirement  obtained from %
	\begin{align}\label{T_eff}
		T^{\mathrm{eff}} = T^{\mathrm{th}} - \max_n(\tau_n).
	\end{align} Given the complexity of jointly selecting relays and minimizing transmission energy, our approach first identifies the optimal transmission link and then focuses on minimizing the transmit energy. 
	By assuming uniform power allocation across all SNs, we reformulate the objective to identify the link that minimizes the transmission delay for each 
	SN, thereby reducing the overall delay while satisfying the delay constraint.
	\begin{subequations} \label{P1_1}
		\allowdisplaybreaks
		\begin{align} &\min_{ \mathcal{N}_{\text{1h}},\mathcal{N}_{\text{2h}},\mathcal{K}}  T_{\textrm{1h}} + T_{\textrm{2h}}^{(1)} +  T_{\textrm{2h}}^{(2)}
			, \label{P1_1-a}\\
			\mathrm{s.t.}\ & 
			\eqref{P1-b}.
		\end{align} 
	\end{subequations}
	To address the problem formulated in \eqref{P1_1}, we propose a threshold-based algorithm aimed at minimizing the transmission delay for each SN. If the total delay exceeds the constraint specified in \eqref{P1-b}, the SN with the weakest link is iteratively removed until the constraint is satisfied.
	The core idea involves selecting a set of SNs that can act as potential relays, denoted by $\mathcal{K}$, based on their channel gains to the ES. This is done by applying a channel gain threshold: SNs with gains above this threshold are considered eligible relays. To determine the optimal threshold within the range bounded by the minimum and maximum channel gains, we employ a ternary search algorithm \cite{TernarySearch1}. This method significantly reduces the computational complexity compared to exhaustive search, and is well-suited for optimizing unimodal, derivative-free functions such as the total system delay.
	For each candidate threshold, SNs are classified into either single-hop or two-hop transmission modes. Specifically, for each SN, we define a delay vector:
	\begin{align}\label{t_n}
		\mathbf{t}_n \triangleq  \left[ \frac{1}{r^{(1)}_{n,1}} + \frac{1}{r^{(2)}_1}, \frac{1}{r^{(1)}_{n,2}} + \frac{1}{r^{(2)}_2}, \ldots, \frac{1}{r^{(1)}_{n,K}} + \frac{1}{r^{(2)}_K} \right].
	\end{align}
	which captures the total transmission time from $\text{S}_n$ to the ES via all possible relays. The ES compares these values against the direct transmission time $1/r^d_n$ and selects the path—either direct or relayed—that yields the minimum delay.
	The full procedure is detailed in Algorithm~\ref{Alg1}.
	To satisfy the delay constraint in \eqref{P1-b}, and inspired by \cite{li2019convergence}, as we will show in Section \ref{Convergence}, where it is demonstrated that increasing the number of successfully participating SNs enhances convergence in FL, we aim to maximize the number of participating SNs.
	This is achieved by iteratively eliminating SNs with the weakest links until the delay constraint \eqref{P1-b} is met.
	The algorithm outputs the indices of SNs categorized into single-hop and two-hop groups, along with the indices of the strongest relays for each SN. To provide a clearer understanding of the classification process, the proposed algorithm is detailed in Algorithm \ref{Alg2}.
	
	\begin{algorithm}[t]
		\caption{SN Classification and Relay Selection via Ternary Search}
		\label{Alg1}
		\KwIn{$h_n^\text{a}$, Rates $r^d, r^{(1)}, r^{(2)}$ with uniform power distribution for $N$ SNs; minimum threshold $th_{\min}$, maximum threshold $th_{\max}$ for channel gain; tolerance $\varepsilon$}
		\KwOut{$\mathcal{N}_{\text{1h}}^*, \mathcal{N}_{\text{2h}}^*, \mathcal{K}^*, \mathbf{T}_d^*$}
		
		\Begin{
			\While{$th_{\max} - th_{\min} > \varepsilon$}{
				$m_1 \gets th_{\min} + \frac{th_{\max} - th_{\min}}{3}$\;
				$m_2 \gets th_{\max} - \frac{th_{\max} - th_{\min}}{3}$\;
				
				Evaluate $T^{\mathrm{UL}}(m_1)$ and $T^{\mathrm{UL}}(m_2)$ using the logic below:\;
				
				\ForEach{$th \in \{m_1, m_2\}$}{
					$\mathcal{N} \gets \{1,\dots,N\}$, $\mathcal{K} \gets \emptyset$, $\mathcal{N}_{\text{1h}} \gets \emptyset$, $\mathcal{N}_{\text{2h}} \gets \emptyset$\;
					\For{$n \leftarrow 1$ \KwTo $N$}{
						\If{$|h_n^\text{a}|^2 > th$}{
							$\text{S}_n \rightarrow \mathcal{K}$, $\mathcal{N} \gets \mathcal{N} \setminus \text{S}_n$\;
						}
					}
					
					\For{$n \leftarrow 1$ \KwTo $N$}{
						Compute $\mathbf{t}_n$ from Eq.~\eqref{t_n}; $j \gets \arg \min(\mathbf{t}_n)$\;
						\If{$\frac{1}{r^d_n} \le \mathbf{t}_n(j)$}{
							$\text{S}_n \rightarrow \mathcal{N}_{\text{1h}}$, $\mathbf{T}_d(n) \gets \frac{1}{r^d_n}$\;
						}
						\Else{
							$\text{S}_n \rightarrow \mathcal{N}_{\text{2h}}$, $\textrm{R}_j \rightarrow \textrm{R}_{k^*_n}$, $\mathbf{T}_d(n) \gets \mathbf{t}_n(j)$\;
						}
					}
					Compute $T^{\textrm{UL}}(th)$ from Eq.~\eqref{eq18}\;
				}
				
				\If{$T^{\mathrm{UL}}(m_1) > T^{\mathrm{UL}}(m_2)$}{
					$th_{\min} \gets m_1$\;
				}
				\Else{
					$th_{\max} \gets m_2$\;
				}
			}
			$th^* \gets \frac{th_{\min} + th_{\max}}{2}$\;
			Evaluate threshold $th^*$ to get $\mathcal{N}_{\text{1h}}^*, \mathcal{N}_{\text{2h}}^*, \mathcal{K}^*, \mathbf{T}_d^*$\;
			\textbf{return} $\mathcal{N}_{\text{1h}}^*, \mathcal{N}_{\text{2h}}^*, \mathcal{K}^*, \mathbf{T}_d^*$
		}
	\end{algorithm}

	\begin{algorithm}[t]
		\caption{Identifying Effective SNs for FL Participation}
		\label{Alg2}
		\KwIn{$\mathcal{N}_{\text{1h}}^*, \mathcal{N}_{\text{2h}}^*, \mathcal{K}^*, \mathbf{T}_d^*$ from Algorithm \ref{Alg1}, $T^\mathrm{eff}$.}
		\KwOut{$\mathcal{N}_{\text{1h}}^*, \mathcal{N}_{\text{2h}}^*, \mathcal{K}^*$.}
		
		\Begin{
			\While{$T^\textrm{UL} > T^\mathrm{eff}$}{
				$n^* \leftarrow \arg \max(\mathbf{T}_d^*)$\;
				\If{$\text{S}_{n^*} \in \mathcal{N}_{\text{1h}}$}{
					$\mathcal{N}_{\text{1h}}^* \leftarrow \mathcal{N}_{\text{1h}}^* \setminus \text{S}_{n^*}$\;
				}
				\Else{
					$\mathcal{N}_{\text{2h}}^* \leftarrow \mathcal{N}_{\text{2h}}^* \setminus \text{S}_{n^*}$\;
				}
			}
			
			\textbf{return} $\mathcal{N}_{\text{1h}}^*, \mathcal{N}_{\text{2h}}^*, \mathcal{K}^*$\;
		}
		
	\end{algorithm}
	Given the sets $\mathcal{N}_{\text{1h}}$, $\mathcal{N}_{\text{2h}}$, and $\mathcal{K}$, we proceed to EE problem. This is achieved by simplifying the optimization problem outlined in \eqref{P1} as follows
	\begin{subequations} \label{P2}
		\allowdisplaybreaks
		\begin{align} &\min_{\mathbf{P}} \left [\sum_{n \in \mathcal{N}_{\text{1h}}} \dfrac{P_n}{ r_{n}^\text{d} }+\sum_{n \in \mathcal{N}_{\text{2h}}}\dfrac{P_n }{   r_{n,k^*_n}^{(1)} }
			+ \sum_{k \in \mathcal{K}} \dfrac{P_{k}}{ r_k^{(2)}}\right], \label{P2-a}\\
			\mathrm{s.t.}\ & 
			\sum_{n \in \mathcal{N}_{\text{1h}}} \dfrac{|B|}{W	 \mathrm{log}_2 \left( 1+ {g}_{n}^d  \right)} \label{P2-b}+ \sum_{n \in \mathcal{N}_{\text{2h}}} \dfrac{|B|}{W  \displaystyle \mathrm{log}_2 \left( 1+ {g}_{n,k^*_n}  \right) }\nonumber\\ & +\sum_{k \in \mathcal{K}} 
			\dfrac{|B|}{W  \mathrm{log}_2 \left( 1+  {g}_{k}  \right)}\leq  T^{\mathrm{eff}}, 
			\\&
			\eqref{P1-c},	
		\end{align} 
	\end{subequations}
	where \eqref{P2-b} refers to the total time constraint. 
	Problem~\eqref{P2} is challenging to solve because of the nonconvex latency constraint~\eqref{P2-b} and the presence of continuous power variables, making the search for a global optimum generally intractable. 
	To overcome this, we adopt SPCA method, which iteratively approximates the original nonconvex problem by a sequence of convex subproblems. 
	At each iteration, the nonconvex constraint is replaced with a locally tight convex surrogate, and the resulting solution sequence is guaranteed to converge to a Karush--Kuhn--Tucker (KKT) stationary point of the original problem.
	Consequently, we can express \eqref{P2} as follows
	\begin{subequations} \label{P3}
		\allowdisplaybreaks
		\begin{align} &\min_{\mathbf{P},\boldsymbol{\omega},\mathbf{q}} \ 
			E_q
			\label{P3-a}\\&
			\mathrm{s.t.}\
			\left [\sum_{n \in \mathcal{N}_{\text{1h}}}  \dfrac{1}{q_n^{(1)}} +\sum_{n \in \mathcal{N}_{\text{2h}}}\dfrac{1}{q_n^{(2)}} +\sum_{k \in \mathcal{K}} \dfrac{1}{q_k^{(3)}} \right] \leq E_q,  \label{P3-a1}\\ & 
			\dfrac{(\omega_n^d)^2}{P_n} \geq q_n^{(1)}, \ \forall  n \in \mathcal{N}_{\text{1h}},
			\label{P3-b}\\&
			\gamma_{n}^d \geq (\omega_n^d)^2, \ \forall  n \in \mathcal{N}_{\text{1h}},
			\label{P3-c}\\&
			\dfrac{(\omega_n^{(1)})^2}{P_n} \geq q_n^{(2)}, \ \forall  n \in \mathcal{N}_{\text{2h}},
			\label{P3-d}\\&
			\gamma_{n,k^*_n}^{(1)} \geq (\omega_n^{(1)})^2, \ \forall  n \in \mathcal{N}_{\text{2h}},
			\label{P3-e}\\&
			\dfrac{(\omega_k^{(2)})^2}{P_{k}} \geq q_k^{(3)}, \ \forall  k \in \mathcal{K},
			\label{P3-f}\\&
			\gamma_{k}^{(2)} \geq (\omega_k^{(2)})^2, \ \forall  k \in \mathcal{K},
			\label{P3-g}\\&
			\sum_{n \in \mathcal{N}_{\text{1h}}} \dfrac{1}{\gamma_n^d} +\sum_{n \in \mathcal{N}_{\text{2h}}}   \dfrac{1}{\gamma_{n,k^*_n}^{(1)}}+ \sum_{k \in \mathcal{K}} \dfrac{1}{ \gamma_{k}^{(2)}} 
			\leq  \dfrac{T^{\mathrm{eff}} W}{|B|},\label{P3-h} 
			\\&
			\mathrm{log}_2 \left( 1+ g_{n}^\text{d}  \right) \geq \gamma_{n}^d, \ \ \forall  n \in \mathcal{N}_{\text{1h}},\label{P3-i}
			\\&
			\mathrm{log}_2 \left( 1+ g_{n,k^*_n}  \right) \geq \gamma_{n,k^*_n}^{(1)}, \  \forall  n \in \mathcal{N}_{\text{2h}},   \label{P3-j}
			\\&
			\mathrm{log}_2 \left( 1+  g_{k}  \right) \geq \gamma_{k}^{(2)}, \   \forall  k \in \mathcal{K},\label{P3-k}
			\\&
			\eqref{P1-c},	
		\end{align} 
	\end{subequations}
	where  $E_q$ is the EE metric, $q_n^{(1)}$, $q_n^{(2)}$, $q_k^{(3)}$, $\omega_n^{d}$, $\omega_n^{(1)}$, $\omega_n^{(2)}$, $\gamma_n^d$, $\gamma_{n,k^*_n}^{(1)}$, and $\gamma_{k}^{(2)}$ are auxiliary variables to approximate the non-convex terms with convex counterparts. 
	It can be perceived that $\gamma_n^d$, $\gamma_{n,k^*_n}^{(1)}$, and $\gamma_{k}^{(2)}$ play the roles of lower bounds for $\mathrm{log}_2 \left( 1+ {g}_{n}^d  \right)$, $\mathrm{log}_2 \left( 1+ {g}_{n,k^*_n}  \right)$, and $\mathrm{log}_2 \left( 1+  g_{k}    \right)$, respectively.
	Increasing the lower-bound values and simultaneously
	reducing the upper bounds will boost the left side of
	the constraints, which are needed here, so that the constraints
	\eqref{P3-h}-\eqref{P3-k} would be active at the optimum.
	The \eqref{P3-h} is convex since it is a linear combination of three quadratic terms over linear functions that is convex \cite{CVX}. 
	The left side of \eqref{P3-b}, \eqref{P3-d}, and \eqref{P3-f} are noncovex. To get rid of nonconvexity, we define the function $\Omega^{[i]}(\omega,z) $, as the
	first-order lower approximation of them as follows
	\begin{align}\label{omega_approx}
		\dfrac{\omega^2}{z}\geq \dfrac{2\omega^{[i]}}{z^{[i]}}\omega - (\dfrac{\omega^{[i]}}{z^{[i]}})^2 z \triangleq
		\Omega^{[i]}(\omega,z),
	\end{align}
	where $(\omega^{[i]},z^{[i]})$ are the values of the variables $(\omega, z)$ at the
	output of the $i$th iteration.
	Affine approximations of constraints \eqref{P3-i}-\eqref{P3-k},  are given by
	\begin{subequations}\label{lem1-formul}
		\allowdisplaybreaks
		\begin{align}
			&1 + \rho_{n} - 2^ {\gamma_n^d}  \geq 0,\ \forall  n \in \mathcal{N}_{\text{1h}},
			\\&
			\rho_{n} \leq \dfrac{P_n|h_n^\text{a}|^2}{\sigma_0},\ \forall  n \in \mathcal{N}_{\text{1h}},
			\\&
			1 + \psi_{n} - 2^ {\gamma_{n,k^*_n}^{(1)} }  \geq 0,\ \forall  n \in \mathcal{N}_{\text{2h}}, 
			\\&
			\psi_{n} \leq \dfrac{P_{n}
				\displaystyle |h_{n,k^*_n}^r |^2}{\sigma_0},\ \forall  n \in \mathcal{N}_{\text{2h}} \\&
			1 + \zeta_{k} - 2^ {\gamma_{K}^{(2)} }  \geq 0,\ \forall  k \in \mathcal{K}, 
			\\&
			\zeta_{k} \leq  \dfrac{P_{k}|  h_k^\text{a}|^2}{\sigma_0},  \forall k \in \mathcal{K},
		\end{align} 
	\end{subequations}
	where $\rho_{n}$, $\psi_{n}$, and $\zeta_{k}$, are auxiliary variables.
	Thus, by replacing constraints \eqref{P3-i}--\eqref{P3-k} with \eqref{lem1-formul}, and using $\Omega^{[i]}(\omega,z)$ for approximating the left side of \eqref{P3-b}, \eqref{P3-d}, and \eqref{P3-f}, the optimization problem \eqref{P3} is rewritten as
	\begin{subequations} \label{P4}
		\allowdisplaybreaks
		\begin{align} 	&\min_{\mathbf{P},\boldsymbol{\omega},\mathbf{q}} \ 
			E_q
			\label{P4-a}\\&
			\Omega^{[i]}(\omega_n^d,P_n)    \geq q_n^{(1)}, \ \forall  n \in \mathcal{N}_{\text{1h}},
			\label{P4-b}\\&          
			\Omega^{[i]}(\omega_n^{(1)},P_n)   \geq q_n^{(2)}, \ \forall  n \in \mathcal{N}_{\text{2h}},
			\label{P4-c}\\&
			\Omega^{[i]}(\omega_k^{(2)},P_{k})     \geq q_k^{(3)}, \ \forall  k \in \mathcal{K},
			\label{P4-d}\\ & 
			\eqref{P1-c},\eqref{P3-a1},\eqref{P3-c},\eqref{P3-e},\eqref{P3-g},\eqref{P3-h},\eqref{lem1-formul}.
		\end{align} 
	\end{subequations}
	The problem \eqref{P4} is a standard convex semidefinite programming (SDP). This can be solved using numerical solvers, such as the SDP tool in CVX \cite{CVX}.
	The SPCA-based algorithm is
	outlined in Algorithm \eqref{Alg2}. In each iteration, problem \eqref{P4} is
	solved and $\omega^{[i]}$
	is updated using the
	corresponding optimized variable. $\epsilon_I$, and $I_{max}$ are the accuracy and the
	maximum number of iterations of the
	algorithm.

	\begin{algorithm}[t]
		\caption{SPCA-based Algorithm for EE}
		\label{Alg3}
		\KwIn{Threshold accuracy \(\epsilon_I\), maximum iterations \(I_{\max}\).}
		\KwOut{\(\mathbf{P}^*\).}
		
		\Begin{
			\textbf{Initialization:} Initialize \(\Omega^{[i]}(\omega_n^d, P_n)\), 
			\(\Omega^{[i]}(\omega_n^{(1)},P_n)\), and
			\(\Omega^{[i]}(\omega_n^{(2)},P_{k^*_n}^s)\) for \(i = 0\)\;
			
			\While{\(\bigl|E_q^{[i+1]} - E_q^{[i]}\bigr| \,\ge \,\epsilon_I\) \textbf{ or } \(i \le I_{\max}\)}{
				\textbf{I:} Find \(E_q^{[i+1]}\) by solving \(\text{(P4)}\)\;
				\textbf{II:} Update the variables 
				\(\boldsymbol{\omega}^{[i+1]} \) and 
				\(\mathbf{P}^{[i+1]} \)\;
				\textbf{III:} \( i \leftarrow i + 1\)\;
			}
			
			\textbf{return} \(\mathbf{P}^*\)\;
		}
	\end{algorithm}

		\subsection{Convergence and Complexity Analysis}
		\subsubsection{Convergence Analysis}
		At each SPCA iteration, the nonconvex objective and constraints in \eqref{P1} 
		are replaced with first-order concave lower bounds, producing a sequence of 
		convex subproblems of the form \eqref{P4}.  
		Since these surrogates are tight at the current operating point, the solution 
		at iteration $i$ remains feasible at iteration $i{+}1$.  
		This ensures a monotonically nonincreasing objective sequence:
		$E_q^{[i+1]} \le E_q^{[i]}
		$.
		Because the transmit-power constraints in \eqref{P1-c} bound the feasible 
		objective, the sequence converges.  
		Due to the local nature of the linearizations in 
		(\ref{P3-b})–(\ref{P3-f}), the algorithm converges to a KKT-compliant 
		stationary point of the original problem, but global optimality cannot be 
		guaranteed.
		
		\subsubsection{Complexity Analysis}
		The ternary-search–based classification requires evaluating  
		$T^{\mathrm{UL}}(th)$ for two candidate thresholds per iteration. 
		Each evaluation involves computing per-SN service times and potential relay 
		assignments, with an overall cost of $\mathcal{O}(N^{2})$.  
		As ternary search needs only $\mathcal{O}(\log(1/\varepsilon))$ iterations, 
		its total runtime remains very small.  
		Thus, Algorithm~\ref{Alg1} provides a globally optimal threshold with negligible
		computational overhead.
		In Algorithm~\ref{Alg3}
		the main computational burden stems from solving the convex subproblem 
		\eqref{P4} in each SPCA iteration.  
		Using interior-point methods, its worst-case complexity is \cite{Boyd}:
		$\mathcal{O}\!\left(\left(\frac{\mathcal{L}-1}{2}\right)^{3.5}
		\log\!\frac{1}{\epsilon_I}\right)$,
		where $\mathcal{L}$ denotes the number of optimization variables in \eqref{P4} and 
		$\epsilon_I$ is the solver accuracy.  
		Since this convex program must be solved repeatedly until SPCA converges, it 
		dominates the overall computational effort.
		
		The total complexity of the proposed framework is governed almost entirely by 
		the SPCA stage, while Algorithm~\ref{Alg1} introduces no significant overhead.

	\section{Convergence Analysis of Our FL Method}
	\label{Convergence}
	
	The convergence of the relay-assisted FL framework is analyzed in this section, highlighting its benefits compared to the standard single-hop FL approach, especially under non-IID data distributions. The analysis is grounded in standard assumptions widely used in the literature \cite{li2019convergence, mcmahan2017communication}, and demonstrates how the relay-assisted approach improves convergence.
	\subsection{Assumptions}
	We adopt the following assumptions for the convergence analysis \cite{li2019convergence}:
	\begin{enumerate}
		\item Each local objective function $F_n(w)$ is $L$-smooth, i.e., for all $w$ and $w'$,
		\begin{equation}
			F_n(w') \leq F_n(w) + \langle \nabla F_n(w), w' - w \rangle + \frac{L}{2} \|w' - w\|^2. \label{eq:smoothness}
		\end{equation}
		
		\item Each $F_n(w)$ is $\mu$-strongly convex, i.e., for all $w$ and $w'$,
		\begin{equation}
			F_n(w') \geq F_n(w) + \langle \nabla F_n(w), w' - w \rangle + \frac{\mu}{2} \|w' - w\|^2. \label{eq:convexity}
		\end{equation}
		
		\item Let $\xi_k$ be sampled from the $n$-th SN’s local data ($D_n$) uniformly at random. The variance of  $S_n$ is bounded for all $n$ by
		\begin{equation}
			\mathbb{E}\big[\|\nabla F_n(w; \xi_k) - \nabla F_n(w)\|^2\big] \leq \delta_n^2, \label{eq:variance}
		\end{equation}
		where $\delta_n$ is defined as the bounded variance of the stochastic gradient estimate at $S_n$.
		\item For all SNs, the expected second-order moment of the norm of the stochastic gradient is uniformly bounded by
		\begin{equation}
			\mathbb{E}\big[\|\nabla F_n(w; \xi_k)\|^2\big] \leq G^2. \label{eq:bounded_moment}
		\end{equation}
	\end{enumerate}
	In addition to the above assumptions, we use the following term
	\begin{align}
		\Gamma = F^* - \sum_{n=1}^N p_n F_n^*, \label{eq:Gamma}
	\end{align}
	to quantify the degree of non-i.i.d, where $F^*$ and $F_n^*$ are the minimum values of $F$ and $F_n$, respectively, 
	and $p_n = \frac{|D_n|}{\sum_{j=1}^N |D_j|}$ is the weight of the $k$-th SN, proportional to its dataset size \cite{mcmahan2017communication}. From $\Gamma$'s definition, the data distribution is i.i.d if $\Gamma = 0$, or non-i.i.d otherwise.
	\subsection{Convergence Bounds} \label{Converg_sec}
	
	\paragraph{Single-Hop FL}
	In single-hop FL, all SNs communicate their local updates directly to the ES. The global objective function is \cite{li2019convergence}
	\begin{equation}
		F(w) = \sum_{k=1}^N p_k F_k(w).\label{eq:global_objective_singlehop}
	\end{equation}
	The convergence bound for FedAvg in single-hop FL, according to \cite{li2019convergence}, is given by
	\begin{align}
		&\mathbb{E}[F(w^U)] - F^* \nonumber \\& \leq \frac{\kappa}{\nu + U - 1} \left( \frac{2(B + C)}{\mu} + \frac{\mu \nu}{2} \mathbb{E}[\|w^0 - w^*\|^2] \right), \label{eq:singlehop_bound}
	\end{align}
	where  $U$ denotes the total number of SGD updates performed by each SN, and $\kappa = \frac{L}{\mu}$ is the condition number. $\nu = \max\{8\kappa, e\}$ where $e$ is the number of local iterations of SGD performed in a SN between two communications,  $w^0$ is the initial value
	of the global model weights and
	\begin{align}
		B &= \sum_{k=1}^N p_n^2 \delta_n^2 + 6L\Gamma + 8(e - 1)^2 G^2, \label{eq:singlehop_B} \\
		C &= \frac{4}{N_{1h}} e^2 G^2. \label{eq:singlehop_C}
	\end{align}
	
	\paragraph{Relay-Assisted FL}
	In the relay-assisted FL framework, the global objective function is defined according to \eqref{global_model} as follows
	\begin{align}
		F(w) = \frac{\sum_{n \in \mathcal{N}_{\text{1h}}} |D_n| F_n(w) + \sum_{k \in \mathcal{K}} |D_k^\text{r}| F_k^{(r)}(w)}{\sum_{n \in \mathcal{N}_{\text{1h}}} |D_n| + \sum_{k \in \mathcal{K}} |D_k^\text{r}|}, \label{eq:global_objective_relay}
	\end{align}
	where $F_k^{(r)}(w)$ aggregates the objectives of two-hop SNs
	\begin{align}
		F_k^{(r)}(w) = &\frac{\sum_{n \in \mathcal{N}_{\text{2h},k}} |D_n| F_n(w) + |D_k| F_k(w)}{|D_k^\text{r}|}.  \label{eq:relay_objective}   
	\end{align}
	The effective variance and heterogeneity terms for relay-assisted FL are
	\begin{align}
		\delta_{\text{eff}}^2 &= \frac{\sum_{n \in \mathcal{N}_{\text{1h}}} |D_n| \delta_n^2 + \sum_{k \in \mathcal{K}} |D_k^\text{r}| \delta_k^2}{\sum_{n \in \mathcal{N}_{\text{1h}}} |D_n| + \sum_{k \in \mathcal{K}} |D_k^\text{r}|}, \label{eq:effective_variance} \\
		\Gamma_{\text{eff}} &= F^* - \frac{\sum_{n \in \mathcal{N}_{\text{1h}}} |D_n| F_n^* + \sum_{k \in \mathcal{K}} |D_k^\text{r}| F_k^*}{\sum_{n \in \mathcal{N}_{\text{1h}}} |D_n| + \sum_{k \in \mathcal{K}} |D_k^\text{r}|}. \label{eq:effective_heterogeneity}
	\end{align}
	By substituting \eqref{eq:effective_heterogeneity} and \eqref{eq:effective_variance} into \eqref{eq:singlehop_bound}, the convergence bound for relay-assisted FL becomes
	\begin{align}
		&\mathbb{E}[F(w^U)] - F^* \leq \frac{\kappa}{\nu + U - 1} \Bigg( \frac{2}{\mu} \Big( \delta_{\text{eff}}^2 + 6L\Gamma_{\text{eff}} \nonumber\\
		&\quad + 8(e - 1)^2 G^2 + \frac{4}{N} e^2 G^2 \Big) + \frac{\mu \nu}{2} \mathbb{E}[\|w^0 - w^*\|^2] \Bigg). \label{eq:relay_bound_simplified}
	\end{align}
	
	\subsection{Discussion}
	The convergence bound for relay-assisted FL  provides explicit insights into how two-level aggregation improves learning efficiency. We now interpret each key term in the bound and compare it to its counterpart in the single-hop FL framework:
	
	\begin{itemize}
		\item \textbf{Impact of Data Heterogeneity:}
		The term $\Gamma_{\text{eff}}$ in \eqref{eq:effective_heterogeneity}, when compared with $\Gamma$ in \eqref{eq:Gamma}, reflects the reduced data heterogeneity in relay-assisted FL. This reduction arises because data from two-hop SNs is first aggregated at relay nodes, producing more representative and smoother model updates. Typically, this results in $\Gamma_{\text{eff}} < \Gamma$, which lowers the second-order error term in the convergence bound \eqref{eq:relay_bound_simplified}, thereby enhancing convergence in non-IID scenarios.
		\item \textbf{Variance Reduction:}
		The term $\delta_{\text{eff}}^2$ captures the effective gradient variance. Relays aggregate gradients from nearby SNs, smoothing local noise before the global model update. This reduces the stochastic variance term compared to the single-hop case where variance accumulates directly from all nodes. As a result, the first-order convergence error becomes smaller, accelerating learning.
		\item \textbf{Role of Local Iterations:}
		The terms $8(e - 1)^2 G^2$ and $\frac{4}{N} e^2 G^2$ in the convergence bound capture the variance due to local updates and the number of participating SNs. In relay-assisted FL, more SNs  participate in training via relay nodes. This leads to a larger effective $N$ , which helps reduce the term $\frac{4}{N} e^2 G^2$. Additionally, two-hop communication via relays can enable better synchronization and model consistency among neighboring nodes, which helps stabilize the gradient noise, even when the number of local steps $e$ is moderate. This ultimately improves the convergence speed while preserving scalability.
		\item \textbf{Faster Convergence Rate:}
		The convergence bound in \eqref{eq:relay_bound_simplified} is inversely proportional to $\nu + U - 1$, where $U$ is the total number of local SGD updates. Therefore, the convergence rate improves as $U$ increases. However, the per-update convergence performance is also determined by the multiplicative constant in \eqref{eq:relay_bound_simplified}. Relay-assisted FL improves this bound by reducing $\delta_{\text{eff}}^2$ and $\Gamma_{\text{eff}}$, which represent the effective variance and data heterogeneity, respectively. As a result, the upper bound on the optimality gap $\mathbb{E}[F(w^U)] - F^*$ decreases more rapidly with each update, yielding a faster convergence rate compared to the single-hop FL approach—even when $\kappa = \frac{L}{\mu}$ remains the same.
	\end{itemize}

		\section{Discussion on Imperfect CSI}
		\label{ICSI-discuss}
		\subsection{Channel Estimation and ICSI Model}
		To capture the impact of ICSI, channel estimation is performed using pilot training sequences for each SN. 
		Assume that each SN employs $L_p$ pilot symbols, with a total pilot duration of $T_p = L_p T_s$, where $T_s = 1/W$ denotes the symbol period. 
		The total pilot training time for all $N$ SNs is subtracted from the transmission slot $T$, yielding the available uplink data transmission time:
		\begin{equation}
			T^{\mathrm{UL}} \triangleq T - N L_p T_s,
		\end{equation}
		where $T^{\mathrm{UL}}$ denotes the effective uplink data duration.
		During pilot transmission, the ES performs MMSE channel estimation. The MMSE channel estimate of $h_n$ is given by		\begin{equation}\label{MMSE_channel}
			\hat{h}_n  =  h_n-\epsilon_n,
		\end{equation}
		where $\hat{h}_n$ and $\epsilon_n$ denote the estimated channel and estimation error, respectively. 
		We model $h_n$ as a general small-scale fading channel with large-scale gain $\beta_n = \mathbb{E}\{|h_n|^2\}$, capturing path loss, shadowing, and allowing for a possible Rician line-of-sight (LOS) component.
 Under pilot-based
MMSE estimation, the second-order statistics of $\hat{h}_n$ and $\epsilon_n$ are
given by \cite{Khosravirad,Ngo2013}
\begin{align}
    \mathbb{E}\{|\hat{h}_n|^2\} &= \beta_n - \sigma_n^e(L_p), \\
    \mathbb{E}\{|\epsilon_n|^2\} &= \sigma_n^e(L_p),
\end{align}
where  $\sigma_n^e(L_p)$ is the estimation error variance \cite{Khosravirad,Ngo2013}
		\begin{align}\label{eq:mmse}
			\sigma_n^e(L_p) = \frac{\beta_n}{1 + L_p g_n^p \beta_n}.
		\end{align}
		Here, $g_n^p$ denotes the per-symbol pilot SNR excluding large-scale fading. 
		As expected, increasing the pilot length $L_p$ or pilot SNR $g_n^p$ improves estimation accuracy, thereby reducing $\sigma_n^e(L_p)$ and enhancing the quality of ICSI.
		\subsection{Effective SNR under MMSE-Based ICSI}
		Consider the uplink signal model
		\begin{equation}
			y_n = \sqrt{P_n}\,h_n\,x_n + z_n, \qquad z_n \sim \mathcal{CN}(0,\sigma_0),
		\end{equation}
		where $x_n$ is the unit-power transmitted symbol, $z_n$ is the complex Gaussian receiver noise. 
		Using the MMSE decomposition in~\eqref{MMSE_channel} with $\hat{h}_n \perp \epsilon_n$, the received signal can be rewritten as
		\begin{equation}
			y_n = \underbrace{\sqrt{P_n}\,\hat{h}_n\,x_n}_{\text{desired signal}}
			+ \underbrace{\sqrt{P_n}\,\epsilon_n\,x_n}_{\text{self-interference}} 
			+ z_n.
		\end{equation}
		The  instantaneous effective SNR can be written as~\cite{Ngo2013,Ngo}
		\begin{equation}\label{SNR-ICSI}
			\hat{g}_n = \frac{P_n |\hat{h}_n|^2}{P_n \sigma_n^e + \sigma_0}.
		\end{equation}
		
		\subsection{SPCA Reformulation under ICSI}
		
		Under MMSE-based ICSI, the effective uplink SINR in~\eqref{SNR-ICSI} yields an achievable
		rate that takes a \emph{difference-of-logs} (DoL) form:
		\begin{equation}
			r_n(P_n)
			= \log_2\!\bigl(1 + (a_n+b_n)P_n\bigr)
			- \log_2\!\bigl(1 + b_n P_n\bigr),
			\label{eq:dol-rate}
		\end{equation}
		where 
		$a_n=\frac{|\hat h_n|^2}{\sigma_0}$, and 
		$b_n=\frac{\sigma_n^{e}}{\sigma_0}$.
		The coefficient $a_n$ represents the normalized estimated-channel power,
		while $b_n$ captures the normalized self-interference introduced by channel
		uncertainty.  
		When $\sigma_n^{e}=0$,~\eqref{eq:dol-rate} reduces to the perfect-CSI expression
		$r_n(P_n)=\log_2(1+a_nP_n)$, showing that the ICSI formulation is a strict
		generalization.
		
		The function in~\eqref{eq:dol-rate} is nonconvex due to the term 
		$\log_2(1+b_n P_n)$.
		Following the SPCA framework, the first (concave) term is kept exact, while the second
		concave term is replaced by its first-order Taylor approximation at the previous
		iterate $P_n^{(m)}$.  
		This yields the concave lower bound
		\begin{align}
			&r_n(P_n)
			\;\ge\; r_n^{\mathrm{low}}(P_n)
			\triangleq \frac{1}{\ln 2}\log\!\bigl(1+(a_n+b_n)P_n\bigr) \nonumber\\
			& -\frac{1}{\ln 2}\!\left[
			\log\!\bigl(1+b_n P_n^{(m)}\bigr)
			+
			\frac{b_n}{1+b_n P_n^{(m)}}\bigl(P_n-P_n^{(m)}\bigr)
			\right].
			\label{eq:rate-lower-bound}
		\end{align}
		The surrogate $r_n^{\mathrm{low}}(P_n)$ is concave in $P_n$ and thus preserves the
		convexity of each SPCA subproblem.
		Accordingly, the rate constraints in~\eqref{P3-i}–\eqref{P3-k} are replaced by
		\begin{align}
			r_n^{\mathrm{low}}(P_n) &\ge \gamma_n^d,  && n\in\mathcal{N}_{\mathrm{1h}},\\
			r_{n,k_n^*}^{\mathrm{low}}(P_n) &\ge \gamma_{n,k_n^*}^{(1)}, && n\in\mathcal{N}_{\mathrm{2h}},\\
			r_k^{\mathrm{low}}(P_k) &\ge \gamma_{k}^{(2)}, && k\in\mathcal{K}.
			\label{eq:icsi-spca-final}
		\end{align}
		
		This modification is the \emph{only} change required to accommodate ICSI.
		All remaining SPCA steps (objective linearization, auxiliary variables, and feasibility
		updates) remain identical to the perfect-CSI implementation.  
		Furthermore, when $\sigma_n^{e}=0$, the surrogate~\eqref{eq:rate-lower-bound} becomes
		exact, and the method reduces seamlessly to the original perfect-CSI SPCA framework.

	\section{Numerical Results}\label{Simulat}

We consider a factory area of $100 \times 100$ $\rm m^2$ with up to 200 SNs, representing a production module in a smart factory. The SNs and the ES are uniformly distributed within the factory. The key simulation parameters are summarized in Table \ref{Table 1}, unless stated otherwise.
The wireless channels between SNs and the ES, as well as among SNs, follow an independent frequency-flat Rician fading model. The path loss model is based on the factory and open-plan building channel model from \cite{Khosravirad}. Shadow fading is incorporated with a standard deviation of 7 dB, as specified in \cite{3GPP}.

\begin{table}[t]
	\small
	\renewcommand{\arraystretch}{1.3}
	\caption{Federated Learning Simulation Parameters}
	\centering
	\label{Table 1}
	\resizebox{\columnwidth}{!}{
		\begin{tabular}{|c|c|}
			\hline
			\textbf{Parameter}  &  \textbf{Value} \\
			\hline \hline
			Factory area & $100 \times 100 \ \rm m^2$ \\ \hline
			Number of SNs &  Up to 200  \\ \hline
			Transmit data size $|B|$ \cite{Energy1} & 1–20 kbits  \\ \hline
			Local dataset size $D_n$ & 200–400 samples  \\ \hline
			Maximum completion time $T^{\mathrm{th}}$ & 60 s \\ \hline
			UL delay requirement $T^{\mathrm{eff}}$ & 4 ms \\ \hline
			Total bandwidth & 100 MHz \\ \hline
			Carrier frequency & 10 GHz \\ \hline
			Maximum transmission power $P_{\max}$ & 23 dBm \\ \hline
			Power spectral density of AWGN & -174 dBm/Hz  \\ \hline
			Effective switched capacitance $\kappa$ \cite{Energy1} & $10^{-28}$  \\ \hline
			Maximum computation capacity $f^{\max}$ & 2 GHz \\ \hline
			CPU cycles per sample $C_n$ \cite{Energy1} & $[1, 2] \times 10^4$ (cycles/sample) \\ \hline
		\end{tabular}
	}
\end{table}

\subsection{FL Convergence Simulation}\label{converg-sim}
\begin{figure}
	\centering
	\includegraphics[width=1\linewidth]{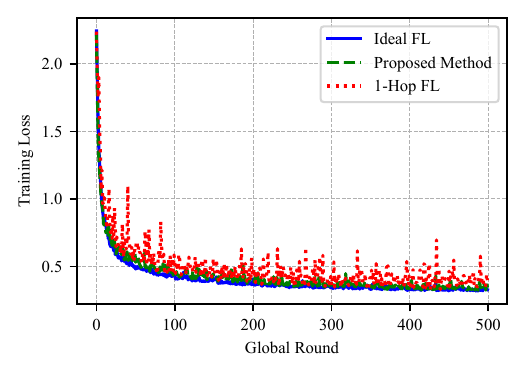}
	\caption{Training loss convergence of FL for different methods over 500 rounds.}
	\label{fig_training}	
\end{figure}

\begin{figure}
	\centering
	\includegraphics[width=1\linewidth]{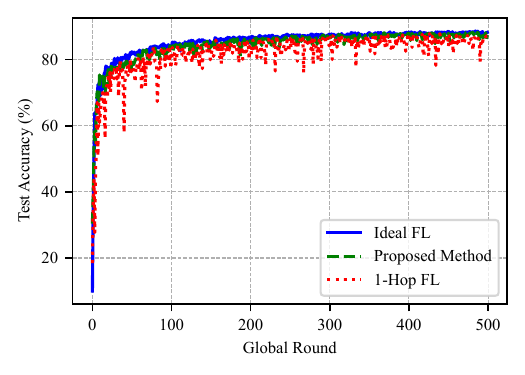}
	\caption{Test accuracy of FL for different methods over 500 rounds.}
	\label{fig_test}	
\end{figure}

To evaluate the performance of the proposed cooperative FL approach, we conduct experiments using the Fashion-MNIST dataset \cite{MNIST} for image classification which serves as a representative benchmark for industrial applications such as visual inspection systems in manufacturing cells.
The dataset consists of 60,000 training samples and 10,000 test samples, spanning 10 fashion categories. Each SN is allocated a random number of training samples following a uniform distribution, $|D_k| \sim U(200, 400)$. In the non-i.i.d. setting, each SN receives data samples corresponding to only two labels.
The learning model is a convolutional neural network (CNN) consisting of two $3\times3$ convolutional layers, each followed by batch normalization and a $2\times2$ max-pooling layer. The model also includes two fully connected layers with a dropout layer in between and a final softmax output layer. Training is performed using the cross-entropy loss function with a learning rate of $\eta = 0.01$, a batch size of $b = 32$, and $e = 3$ local epochs per client.
We compare three FL setups:
\begin{itemize}
	\item \textbf{Ideal FL}: A basic scheme assuming perfect communication, involving 200 SNs with 50 randomly selected SNs participating in each training round.
	\item \textbf{Cooperative FL}: Our proposed method with selected SNs determined by Algorithm \ref{Alg1} and Algorithm \ref{Alg2}.
	\item \textbf{1-Hop FL}: Only single hop transmission with selected SNs determined by Algorithm \ref{Alg2}.
\end{itemize}
All methods use FedAvg over 500 global rounds. Performance is evaluated in terms of training loss and test accuracy across communication rounds, as illustrated in Fig.~\ref{fig_training} and Fig.~\ref{fig_test}. To quantitatively compare the convergence behavior of different schemes, we use the \textit{Normalized Mean Squared Error} (NMSE) metric, defined as
\[
\text{NMSE} = \frac{\sum_{m=1}^{M} (\hat{y}_m - y_m)^2}{\sum_{m=1}^{M} y_m^2},
\]
where \( y_m \) and \( \hat{y}_m \) represent the values from the Ideal FL and the method under evaluation at global round \( m \), respectively, and \( M \) is the total number of rounds.
Table~\ref{tab:nmse_results} reports the NMSE values for both training loss and test accuracy. For training loss, our method achieves an NMSE of 0.006 compared to 0.059 for 1-Hop FL. For test accuracy, the NMSE values are 0.0004 for our method and 0.0027 for 1-Hop FL. These results demonstrate a reduction in NMSE by up to one order of magnitude compared to the 1-Hop FL baseline, indicating more stable and accurate convergence behavior.
These results confirm that the cooperative FL method not only achieves competitive test accuracy and lower training loss compared to 1-Hop FL, but also exhibits convergence behavior that closely aligns with the Ideal FL baseline, as further quantified by the NMSE analysis. This is attributed to the increased number of participating SNs per round, which improves learning stability. 
Specifically, for a maximum transmission power of $P_{\max} = 12$ dBm, the average number of selected SNs for cooperative FL and 1-Hop FL are 28 and 9, respectively, further highlighting the advantage of our method in FL deployment.
\begin{table}[t]
	\centering
	\caption{NMSE Compared to Ideal FL}
	\label{tab:nmse_results}
	\begin{tabular}{lcc}
		\toprule
		\textbf{Scheme} & \textbf{NMSE (Accuracy)} & \textbf{NMSE (Loss)} \\
		\midrule
		Proposed Method & 0.0004 & 0.006 \\
		1-Hop FL        & 0.0027 & 0.059 \\
		\bottomrule
	\end{tabular}
\end{table}


\subsection{Evaluation of Outage and Effective SNs}
In this subsection, we first evaluate the outage probability.
In our simulation, we define an outage as an overflow in the UL transmission time during a round of FL training. We evaluate the performance of Algorithm~\ref{Alg1} by comparing our proposed method against several benchmark schemes\footnote{These benchmarks are designed to highlight the individual contributions of key components in the proposed algorithm.}:
1) \textbf{Alg. 1 (Proposed Method)}: The method based on Algorithm \ref{Alg1}.
2) \textbf{Alg. 1, Fixed \textit{th}}: Algorithm \ref{Alg1} with a fixed \textit{th} instead of searching for the optimal \textit{th}. The average channel gain in each round is used as the threshold.
3) \textbf{Only 2-hop, Fixed \textit{th}}: Assumes all SNs use 2-hop transmission, with relays selected based on the average channel gain.
4) \textbf{Only 1-hop}: Direct single-hop transmission.
5) \textbf{Random Selection}: A relay selection strategy where relays are chosen randomly.

\begin{figure} 
	\includegraphics[width=1\linewidth]{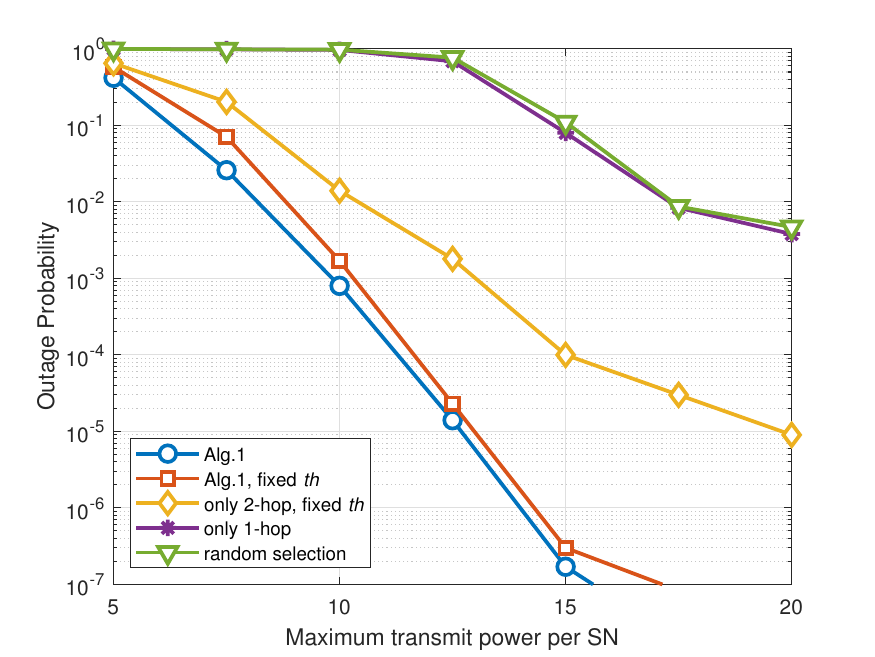}
	\caption{Outage probability versus the maximum transmit power per SN for different methods with  $|B| =1 $ kbits.}
	\label{fig_outage}
\end{figure}

Fig. \ref{fig_outage} presents the outage probability versus the maximum transmit power per SN for these methods. The packet size is set to $|B| = 1$ kbit throughout the simulation.
It is evident that our proposed method outperforms all benchmarks. The adaptive threshold approach slightly reduces the outage compared to using a fixed threshold. Solely relying on 2-hop transmission results in a higher outage, as in some cases, single-hop transmission offers lower delay due to shorter distances to the ES and better channel conditions. Finally, random relay selection and single-hop transmission exhibit similar performance, as leveraging random relays does not necessarily provide the benefits of lower delay.

\begin{figure}
	\centering
	\includegraphics[width=1\linewidth]{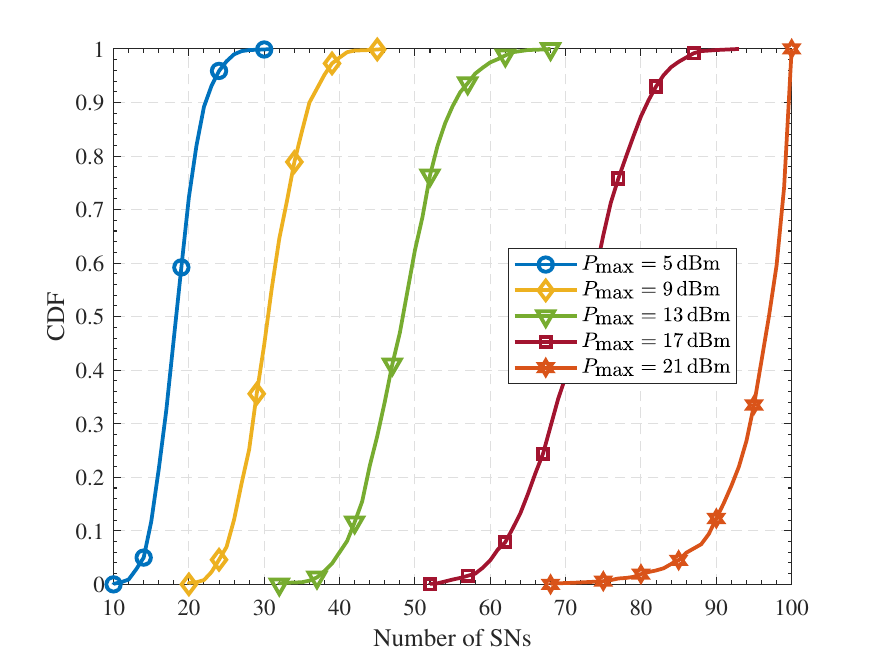}
	\caption{Empirical CDF of the number of SNs participating in each round of FL for different values of $P_{\max}$ using the proposed algorithm.}
	\label{figCDF}	
\end{figure}

To examine Algorithm \ref{Alg2}, we plot the empirical cumulative distribution function (CDF) of the number of SNs participating in each round of FL for different values of $P_{\max}$ in Fig. \ref{figCDF}. It is evident that increasing the transmit power allows more SNs to be selected for FL. Specifically, when $P_{\max}=21$ dBm, in most cases, more than 90 out of 100 SNs can satisfy the delay constraint.
It is also worth noting that relaxing the delay constraint allows more SNs with lower transmission power to participate in the FL process. This increased participation can improve model diversity and accelerate convergence, as supported by the analysis in subsection~\ref{converg-sim}.


\subsection{EE Evaluation}  

The objective of Algorithm \ref{Alg3} is to optimize power allocation in order to minimize energy consumption. In this subsection, we compare the proposed method with four baseline schemes:  
1) 2-Hop without partial aggregation (2-Hop-wo-PA),  
2) 1-Hop,  
3) 1-Hop with maximum transmit power (\(P_{\max}\)), and  
4) RIS-assisted communication with 64 and 128 reflecting elements.

\begin{figure} 	
	\centering
	{\includegraphics[width=.95\linewidth]{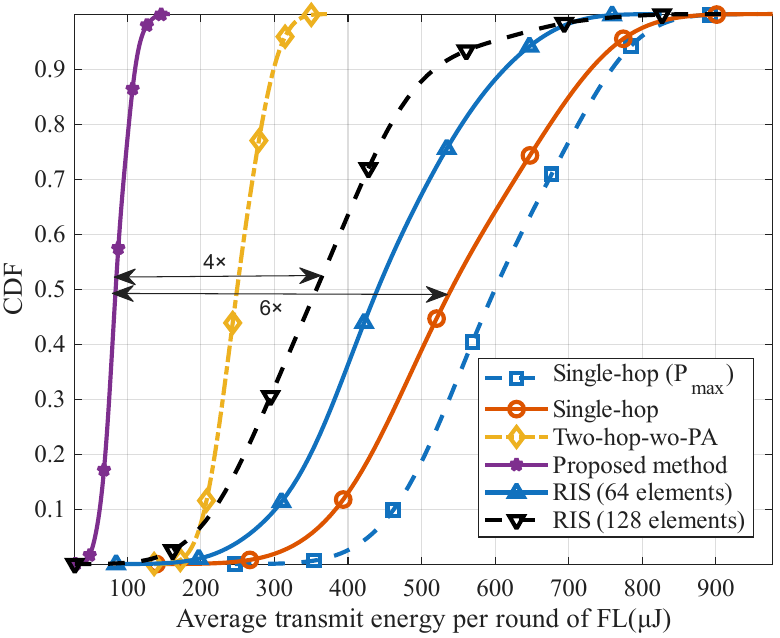}}%
	\caption{CDF of total transmission energy for  50 SNs with a data size of $|B| = 5$ kbit  per  FL round, under different schemes.}
	\label{fig_E_CDF}	
\end{figure}

Fig.~\ref{fig_E_CDF} presents the empirical CDF of the total transmission energy for 50~SNs, each transmitting a data size of \( |B| = 5 \)~kbit per FL round. As observed, the proposed method achieves the best EE among all schemes. Specifically, it reduces communication energy by up to a factor of six compared to the 1-Hop scheme and by nearly a factor of three relative to 2-Hop-wo-PA. 
For the RIS benchmark, the RIS is randomly deployed within a radius of 10 m from the ES, while maintaining a minimum distance of 2 m; each SN communicates via both direct and RIS-reflected links. Due to the adopted TDMA protocol, only one SN is active per time slot, thereby enabling optimal RIS phase configuration via closed-form phase alignment \cite{Hashempour}. In particular, the RIS is configured to achieve constructive signal combining by aligning the phase of the reflected link with that of the direct link at the receiver, thereby enhancing the overall received signal power. The transmit power is then optimized using the same SPCA-based framework as in the single-hop case, as in \cite{Hashempour}. 
As shown in Fig.~\ref{fig_E_CDF}, increasing the number of RIS elements from 64 to 128 improves EE due to enhanced passive beamforming gain. However, even with 128 elements, the proposed relay-assisted scheme achieves up to a factor of four lower energy consumption, demonstrating its clear advantage.

This performance gain stems from the active nature of relaying, which strengthens weak links by splitting long transmission distances into shorter hops. In contrast, RIS-assisted communication is passive and suffers from the multiplicative double path-loss effect of cascaded channels. Moreover, although RIS is assumed to be energy-neutral, the proposed scheme does not require dedicated relay devices and instead reuses existing SNs for cooperation, thereby avoiding additional infrastructure and energy overhead while maintaining high EE.

The performance gap between 2-Hop-wo-PA and the proposed method arises from the absence of partial aggregation. Without aggregation, each SN’s full model must be forwarded individually, leading to increased communication load. In contrast, the proposed scheme aggregates updates at the relay, thereby reducing the total transmitted data and significantly lowering energy consumption.

\begin{figure} 
	\centering
	{\includegraphics[width=1\linewidth]{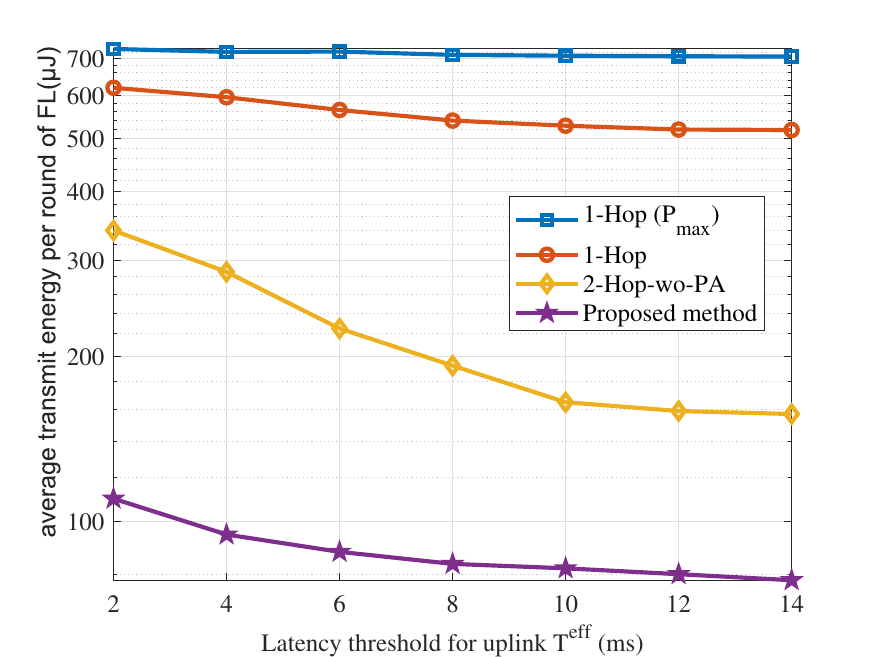}}%
	\caption{Comparison of average transmission energy for 50 SNs, each transmitting $|B| = 5$ kbit per FL round, as a function of the latency threshold under different schemes.}
	\label{fig_E_delay}	
\end{figure}

\begin{figure}
	\centering
	\includegraphics[width=1\linewidth]{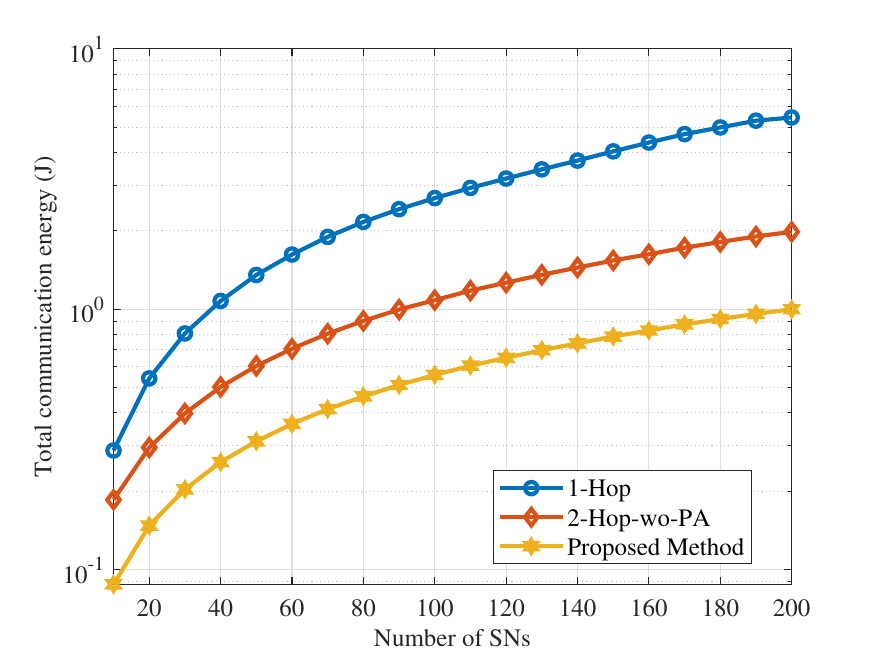}
	\caption{Wireless transmit energy versus number of  SNs for different communication schemes with fixed target accuracy.}
	\label{fig-WTE}	
\end{figure}

In the next simulation, we investigate the impact of the latency threshold (\(T^{\mathrm{eff}}\)) on the average transmission energy per FL round as shown in Fig. \ref{fig_E_delay}. 
 As the latency threshold $T^{\mathrm{eff}}$ increases from $2$~ms to $14$~ms, the average
transmission energy decreases monotonically. A more relaxed latency constraint allows each
SN to transmit its local model over a longer duration using lower transmit power, thereby
reducing the total uplink energy. As $T^{\mathrm{eff}}$ becomes large, the rate of energy
reduction gradually diminishes and the curves approach saturation. By defining  $c_i \triangleq \frac{|h_i|^2}{\sigma_0}$, this behavior can be
explained by the low-SNR approximation
$r_i(P_i) = \log_2\bigl(1 + c_i P_i\bigr) \approx c_i P_i /ln2,$ which implies that the per-SN energy term
$E_i \triangleq \frac{P_i}{r_i(P_i)} \approx ln2/c_i$
becomes nearly constant, providing little additional gain from further relaxing $T^{\mathrm{eff}}$.
Across all latency thresholds, the proposed scheme consistently outperforms the baselines,
achieving up to six-fold energy reduction compared with the 1-Hop scheme and at least a
two-fold improvement over the 2-Hop-wo-PA case.
 While the results are shown for 50 SNs with \( |B| = 5 \)~kbit, the approach generalizes to other values as well, demonstrating robust energy savings. For brevity and to avoid redundancy, we omit additional cases that follow the same trend.
To evaluate the total energy consumption, Fig.~\ref{fig-WTE} illustrates the total communication energy versus the number of SNs for various communication schemes, using the Fashion-MNIST dataset and the previously described settings. As expected, increasing the number of SNs from 10 to 200 leads to a rise in total transmission energy across all schemes. However, our proposed method consistently achieves lower energy consumption—at least two times less than the other approaches—highlighting its scalability and effectiveness in energy-efficient communication, even as the network size grows.

\begin{figure}
	\centering
	\includegraphics[width=1\linewidth]{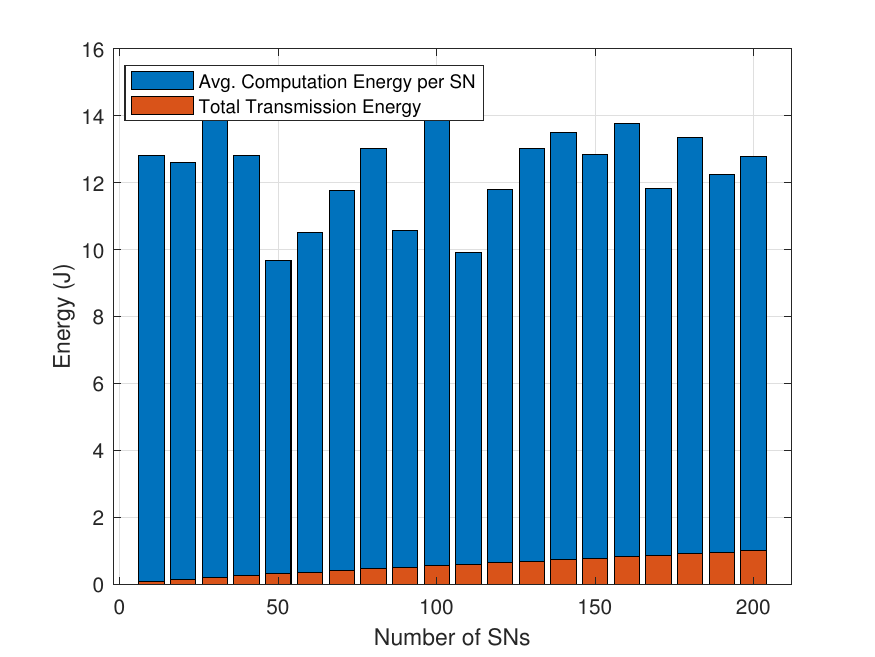}
	\caption{Comparison of total transmission energy and average computation energy per SN as a function of the number of SNs.}
	\label{fig5}	
\end{figure}

Finally, Fig.~\ref{fig5} compares the total transmission energy with the average computation energy per SN. In our specific small-scale factory environment, characterized by the presence of line-of-sight (LOS) links as modeled by 3GPP, and a relatively small packet size of \( |B| = 10 \)~kbit, computation energy significantly outweighs transmission energy. In fact, communication energy accounts for only 5\% to 10\% of the total energy per SN in this scenario. However, in practical deployments where the packet size can reach the order of megabits, the wireless transmission energy could become comparable to or even exceed computation energy by several orders of magnitude. It is important to note that our proposed method achieves up to a 6-fold reduction in communication energy compared to the 1-Hop baseline. This implies that our method can reduce communication energy from a potential 60\% of the computation energy down to under 10\%, thereby delivering substantial energy savings. Moreover, in more challenging environments with non-line-of-sight (NLOS) conditions and larger packet sizes, the advantages of our approach become even more pronounced, underscoring its effectiveness in energy-efficient FL.

	\subsection{Effect of Imperfect CSI}
	
\begin{figure}
		\centering
		\includegraphics[width=1\linewidth]{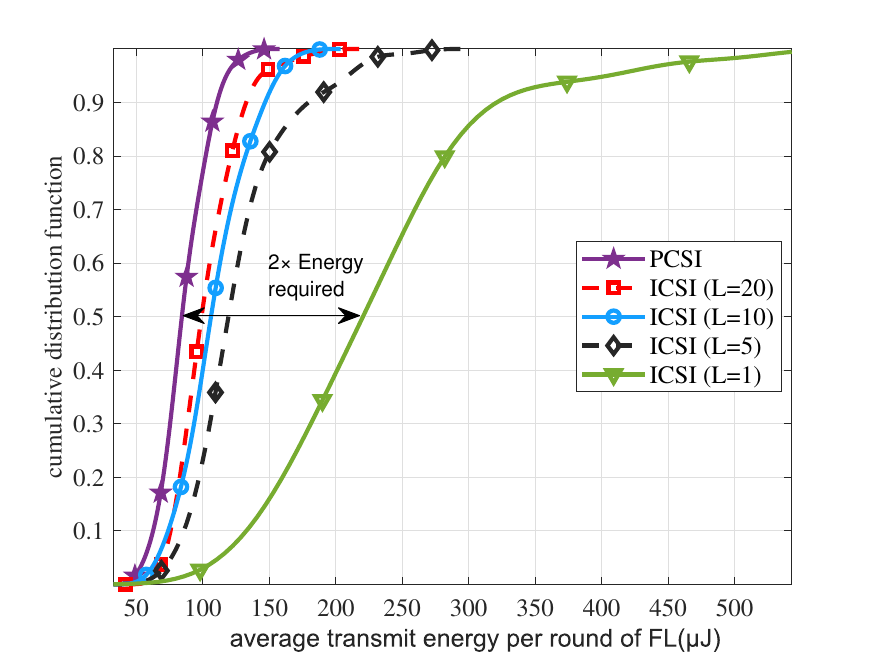}
		\caption{CDF of uplink transmit energy for 50 SNs with a data size of $|B|=5$ kbit per FL round under different CSI conditions.}
		\label{fig:cdf_icsi_pcsi}
	\end{figure}

	Fig.~\ref{fig:cdf_icsi_pcsi} shows the CDF of the average uplink transmit energy under PCSI and ICSI with pilot lengths $L_p\!\in\!\{1,5,10,20\}$.  
	Because the effective SINR under ICSI,
	$\hat{g}_n$ (according to \eqref{SNR-ICSI}),
	is always lower than in the PCSI case, the ICSI curves are shifted to the right, indicating higher required transmit energy.
	Despite this, the proposed SPCA method remains robust: all CDFs preserve a similar shape, and the degradation relative to PCSI is limited. Increasing the pilot length $L_p$ reduces the estimation error $\sigma_n^e$, thereby narrowing the gap. For example, with $L_p=1$, the median energy is more than twice that of PCSI, whereas with $L_p=20$, the gap is reduced to roughly $10\,\mu\text{J}$, which is negligible compared to the $L_p=1$ case.
	These results demonstrate that the proposed approach maintains strong energy performance even under practical channel-estimation uncertainty.
	
\section{Conclusion}\label{conc}  
In this paper, we proposed an EE-FL framework where SNs transmit their locally trained models using either single-hop or two-hop communication. To reduce packet size and communication overhead, we introduced partial aggregation at the relay stage, enhancing the overall efficiency of the FL process.
Our goal was to minimize total energy consumption through a joint optimization of system parameters. Given the coupling between variables, we decomposed the problem and adopted a sequential optimization strategy. First, we developed an algorithm for relay selection and effective SN participation per round. Next, we optimized the operating frequency of each SN to reduce computation energy. Finally, we determined the optimal transmit power to minimize communication energy.
The proposed SPCA algorithm successfully balances communication and local computation costs to manage energy consumption efficiently. 
Furthermore, the framework was extended to the ICSI case by incorporating MMSE-based channel estimation and deriving the corresponding effective-SNR model, enabling a unified SPCA solution under both PCSI and ICSI.

Simulation results confirm the effectiveness of our approach, showing improved convergence, a substantial reduction in outage probability (from \(10^{-2}\) in the single-hop case to \(10^{-6}\) with SPCA), and significant energy savings achieving at least a twofold reduction compared to the cooperative scheme without aggregation, and up to six-fold
lower energy consumption than single-hop transmission. These findings demonstrate the scalability and robustness of our method, particularly in communication-constrained environments.

Finally, while the proposed scheme ensures synchronized aggregation by enforcing strict per-round latency constraints, we do not explicitly model stochastic synchronization errors or update staleness. Studying synchronization-aware FL with relay-induced delays constitutes an important extension for future investigation.
Moreover, our primary goal in this paper is EE in FL. Deriving explicit methods for maximizing FL convergence in IIoT networks will be investigated in future work.

\end{document}